\RequirePackage[hyphens]{url}

\documentclass{article}
\usepackage[utf8]{inputenc}
\usepackage{graphicx}
\usepackage{paralist}
\usepackage[export]{adjustbox}
\usepackage{authblk}
\usepackage{csvsimple}
\usepackage{longtable}
\usepackage{threeparttablex}
\usepackage{tabularx}
\usepackage{paralist}
\usepackage{titlesec}
\usepackage{soul}
\usepackage{amssymb}
\usepackage{hyperref}

\newcommand{\mysection}[1]{\section{#1}}

\title{The spread of low-credibility content \\ by social bots}

\author{Chengcheng Shao}
\author{Giovanni Luca Ciampaglia}
\author{Onur Varol}
\author{Kaicheng Yang}
\author{Alessandro Flammini}
\author{Filippo Menczer}

\affil{Indiana University, Bloomington}

\date{}

\begin{document}

\def \bigfig{0.8\textwidth}
\def \mediumfig{0.75\textwidth}
\def \smallfig{0.65\textwidth}
\def \halffig{0.45\textwidth}

\maketitle

\begin{abstract}
    The massive spread of digital misinformation has been identified as a major global risk and has been alleged to influence elections and threaten democracies. Communication, cognitive, social, and computer scientists are engaged in efforts to study the complex causes for the viral diffusion of misinformation online and to develop solutions, while search and social media platforms are beginning to deploy countermeasures. With few exceptions, these efforts have been mainly informed by anecdotal evidence rather than systematic data. 
    Here we analyze 14 million messages spreading 400 thousand articles on Twitter during and following the 2016 U.S. presidential campaign and election. We find evidence that social bots played a disproportionate role in amplifying low-credibility content.
    Accounts that actively spread articles from low-credibility sources are significantly more likely to be bots. 
    Automated accounts are particularly active in amplifying content in the very early spreading moments, before an article goes viral. 
    Bots also target users with many followers through replies and mentions.
    Humans are vulnerable to this manipulation, retweeting bots who post links to low-credibility content. Successful low-credibility sources are heavily supported by social bots. 
    These results suggest that curbing social bots may be an effective strategy for mitigating the spread of online misinformation. 
\end{abstract}

\mysection{Introduction}

If you get your news from social media, as most Americans do~\cite{Pew2016}, you are exposed to a daily dose of false or misleading content --- hoaxes, conspiracy theories, fabricated reports, click-bait headlines, and even satire. We refer to such content collectively as ``misinformation.'' The incentives are well understood: traffic to fake news sites is easily monetized through ads~\cite{Markines09airweb}, but political motives can be equally or more powerful~\cite{mustafaraj2010obscurity,Truthy_icwsm2011class}. The massive spread of digital misinformation has been identified as a major global risk~\cite{Forum:2013}. Claims that so-called ``fake news'' can influence elections and threaten democracies~\cite{TrendMicro}, however, are hard to prove~\cite{allcott2017social}. Yet we have witnessed abundant demonstrations of real harm caused by misinformation and disinformation spreading on social media, from dangerous health decisions~\cite{10.1371/journal.pmed.1002153} to manipulations of the stock market~\cite{socialbots-CACM}.

%%% SHORT VERSION
A complex mix of cognitive, social, and algorithmic biases contribute to our vulnerability to manipulation by online misinformation~\cite{fake-news-manifesto}. 
These include information overload and finite attention~\cite{low_quality_nhb2017}, novelty of false news~\cite{Vosoughi1146}, the selective exposure~\cite{Sunstein_extremes,Pariser,Nikolov15socialbubbles} caused by polarized and segregated online social networks~\cite{conover12partisan,Truthy_icwsm2011politics}, algorithmic popularity bias~\cite{Salganik,Hodas12socialcom,Azadeh_pop_2017}, and other cognitive vulnerabilities such as confirmation bias and motivated reasoning~\cite{Stroud2011,KahanIdeology2013,Levendusky2013}.  
%%% LONG VERSION
%Even in an ideal world where individuals tend to recognize and avoid sharing low-quality information, information overload and finite attention limit the capacity of social media to discriminate information on the basis of quality~\cite{low_quality_nhb2017}. In the real world, attention can easily be tricked by the novelty of misinformation and as a result, false news are as likely to go viral as reliable information, or more~\cite{Vosoughi1146}. Furthermore, our online social networks are strongly polarized and segregated along political lines~\cite{conover12partisan,Truthy_icwsm2011politics}. The resulting ``echo chambers''~\cite{Sunstein_extremes,Pariser} provide selective exposure to news sources, biasing our view of the world~\cite{Nikolov15socialbubbles}. Finally, social media platforms are designed to prioritize engaging posts. Such algorithmic popularity bias may well hinder the selection of quality content~\cite{Salganik,Hodas12socialcom,Azadeh_pop_2017}. All of these factors play into confirmation bias and motivated reasoning~\cite{Stroud2011,KahanIdeology2013,Levendusky2013}, making the truth hard to discern. 

Abuse of online information ecosystems can both exploit and reinforce these vulnerabilites. While fabricated news are not a new phenomenon~\cite{Lippmann1922}, the ease with which social media can be manipulated~\cite{Truthy_icwsm2011class} creates novel challenges and particularly fertile grounds for sowing disinformation. Public opinion can be influenced thanks to the low cost of producing fraudulent websites and high volumes of software-controlled profiles or pages, known as \emph{social bots}~\cite{socialbots-CACM, botornot_icwsm17}. These fake accounts can post content and interact with each other and with legitimate users via social connections, just like real people~\cite{socialbots-IEEE-DARPA}. Bots can tailor misinformation and target those who are most likely to believe it, taking advantage of our tendencies to attend to what appears popular, to trust information in a social setting~\cite{Jun06062017}, and to trust social contacts~\cite{Menczer06socialphishing}. Bots alone may not entirely explain the success of false news, but they do contribute to it~\cite{Vosoughi1146}. Since earliest manifestations uncovered in 2010~\cite{mustafaraj2010obscurity,Truthy_icwsm2011class}, we have seen influential bots affect online debates about vaccination policies~\cite{socialbots-CACM} and participate actively in political campaigns, both in the U.S.~\cite{bessi2016social} and other countries~\cite{Howard2017,ferrara2017frenchbots}. 

The fight against online misinformation requires a grounded assessment of the relative impact of different mechanism by which it spreads. If the problem is mainly driven by cognitive limitations, we need to invest in news literacy education; if social media platforms are fostering the creation of echo chambers, algorithms can be tweaked to broaden exposure to diverse views; and if malicious bots are responsible for many of the falsehoods, we can focus attention on detecting this kind of abuse. Here we focus on gauging the latter effect. Most of the literature about the role played by social bots in the spread of misinformation is based on anecdotal or limited evidence; a quantitative understanding of the effectiveness of misinformation-spreading attacks based on social bots is still missing. A large-scale, systematic analysis of the spread of low-credibility content by social bots is now feasible thanks to two tools developed in our lab: the \emph{Hoaxy} platform to track the online spread of claims~\cite{Shao2018anatomy} and the \emph{Botometer} machine learning algorithm to detect social bots~\cite{botornot_icwsm17}. Here we examine social bots and how they promote this class of content 
%hundreds of thousands of false and misleading articles 
spreading through millions of Twitter posts during and following the 2016 U.S. presidential campaign. 

\mysection{Results}

Our analysis is based on a large corpus of news stories posted on Twitter. Rather than focusing on individual stories that have been flagged as misinformation by fact-checkers, we track the complete production of a number of sources known to fact-checkers for their low credibility.
%Our analysis is based on the spread of content from low-credibility sources rather than focusing on individual stories that are labeled as misinformation. 
There are two reasons for this approach~\cite{fake-news-manifesto}. First, these sources have intent and processes for the deception and manipulation of public opinion. Second, fact-checking millions of individual articles is unfeasible. As a result, this approach is widely adopted in the literature (see Supplementary Materials). 
The links to the articles considered here were crawled from 120 \emph{low-credibility sources} that, according to lists compiled by reputable third-party news and fact-checking organizations, routinely publish various types of false and/or misleading news. Our own analysis of a sample of articles confirms that the vast majority of their content is some type of misinformation (see Methods). 
We also crawled the articles published by seven independent fact-checking organizations for the purpose of comparison.  
The present analysis focuses on the period from mid-May 2016 to the end of March 2017. During this time, we collected 389,569 articles from low-credibility sources and 15,053 articles from fact-checking sources. We further collected from Twitter \emph{all} of the public posts linking to these articles: 13,617,425 tweets linked to low-credibility sources and 1,133,674 linked to fact checking sources. See Methods and Supplementary Materials for details. 

% Repo: https://github.com/IUNetSci/HoaxyBots
% website: https://iunetsci.github.io/HoaxyBots/

\begin{figure}
\centerline{\includegraphics[width=\bigfig]{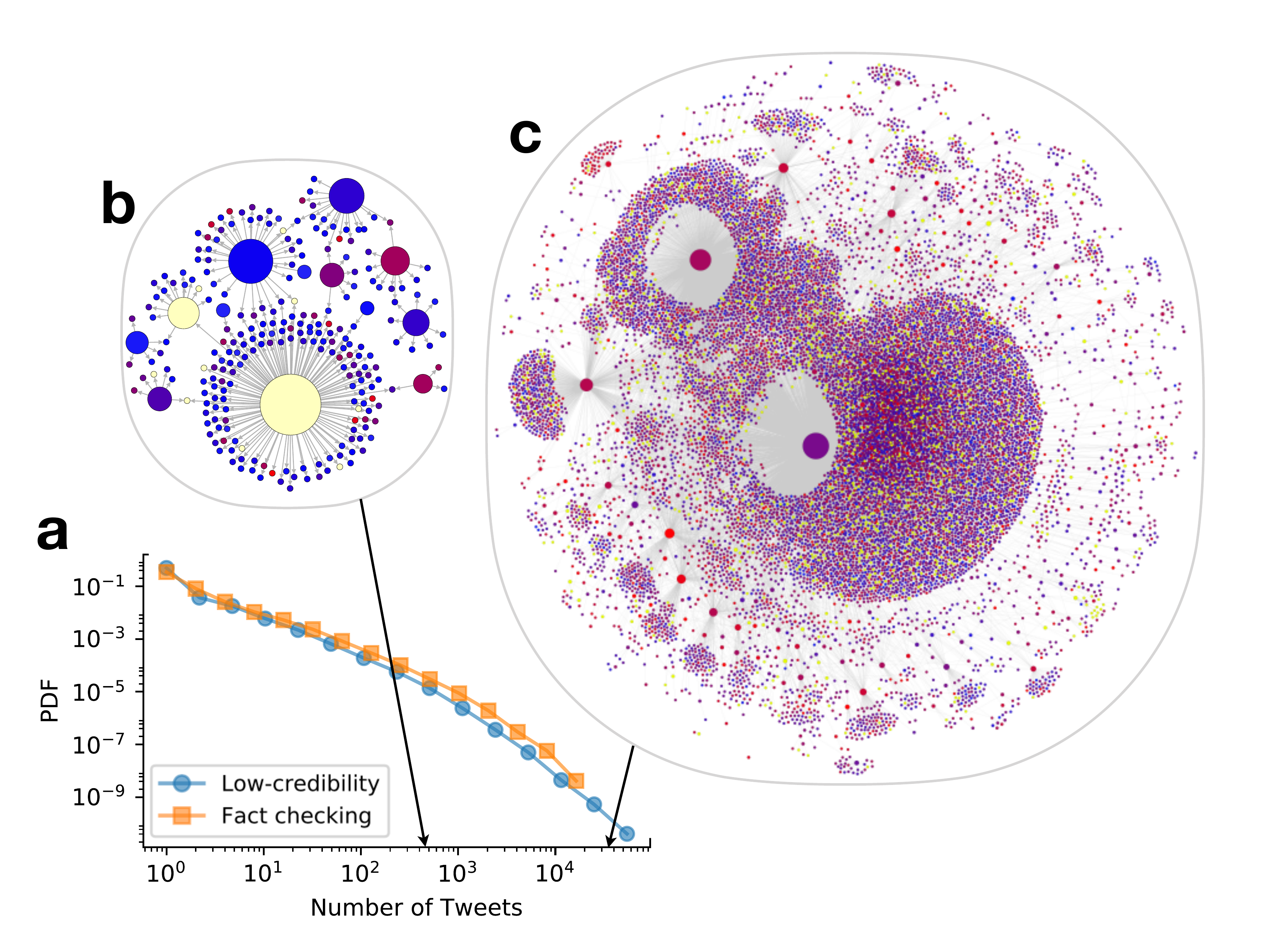}}
\caption{Online virality of content. (a)~Probability distribution (density function) of the number of tweets for articles from both low-credibility and fact-checking sources. The distributions of the number of accounts sharing an article are very similar (see Supplementary Materials). As illustrations, the diffusion networks of two stories are shown: (b)~a medium-virality misleading article titled \textit{FBI just released the Anthony Weiner warrant, and it proves they stole election,} published 
%by the partisan site \protect\url{OccupyDemocrats.com} 
a month after the 2016 U.S. election and shared in over 400 tweets; and (c)~a highly viral fabricated news report titled \textit{``Spirit cooking'': Clinton campaign chairman practices bizarre occult ritual,} published 
%by the conspiracy site \protect\url{Infowars.com} 
four days before the 2016 U.S. election and shared in over 30 thousand tweets. In both cases, only the largest connected component of the network is shown. Nodes and links represent Twitter accounts and retweets of the article, respectively. Node size indicates account influence, measured by the number of times an account was retweeted. Node color represents bot score, from blue (likely human) to red (likely bot); yellow nodes cannot be evaluated because they have either been suspended or deleted all their tweets. An interactive version of the larger network is available online (\protect\url{iunetsci.github.io/HoaxyBots/}).
Note that Twitter does not provide data to reconstruct a retweet tree; all retweets point to the original tweet. The retweet networks shown here combine multiple cascades (each a ``star network'' originating from a different tweet) that all share the same article link.}
\label{fig:a-pdf-tweets-and-users}
\end{figure}

On average a low-credibility source published approximately 100 articles per week. By the end of the study period, the mean popularity of those articles was approximately 30 tweets per article per week (see Supplementary Materials). However, as shown in Fig.~\ref{fig:a-pdf-tweets-and-users}, success is extremely heterogeneous across articles.
Whether we measure success by number of accounts sharing an article, or by number of posts containing a link, we find a very broad distribution of popularity spanning several orders of magnitude: while the majority of articles goes unnoticed, a significant fraction goes ``viral.'' Unfortunately, and consistent with prior analysis of Facebook data~\cite{low_quality_nhb2017}, we observe that the popularity distribution of low-credibility articles is indistinguishable from that of fact-checking articles. This result is not as bleak as that of a similar analysis based on only fact-checked claims, which found false news to be even more viral than real news~\cite{Vosoughi1146}. However, the qualitative conclusion is the same: massive numbers of people are exposed to low-credibility content. 

\begin{figure}
\centering
\begin{tabular}{cc}
{\sffamily\Large a}  \includegraphics[width=\halffig,valign=t]{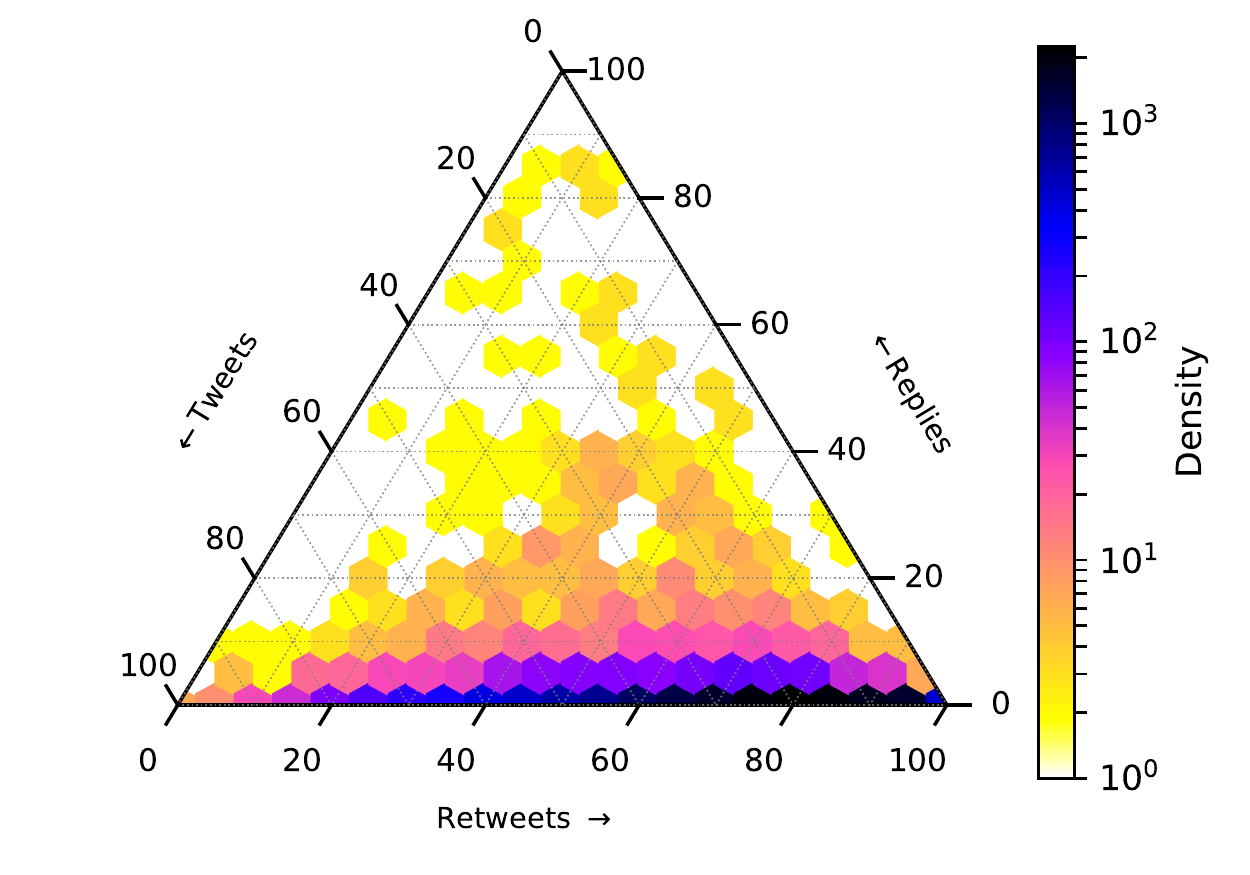} &
{\sffamily\Large b}  \includegraphics[width=\halffig,valign=t]{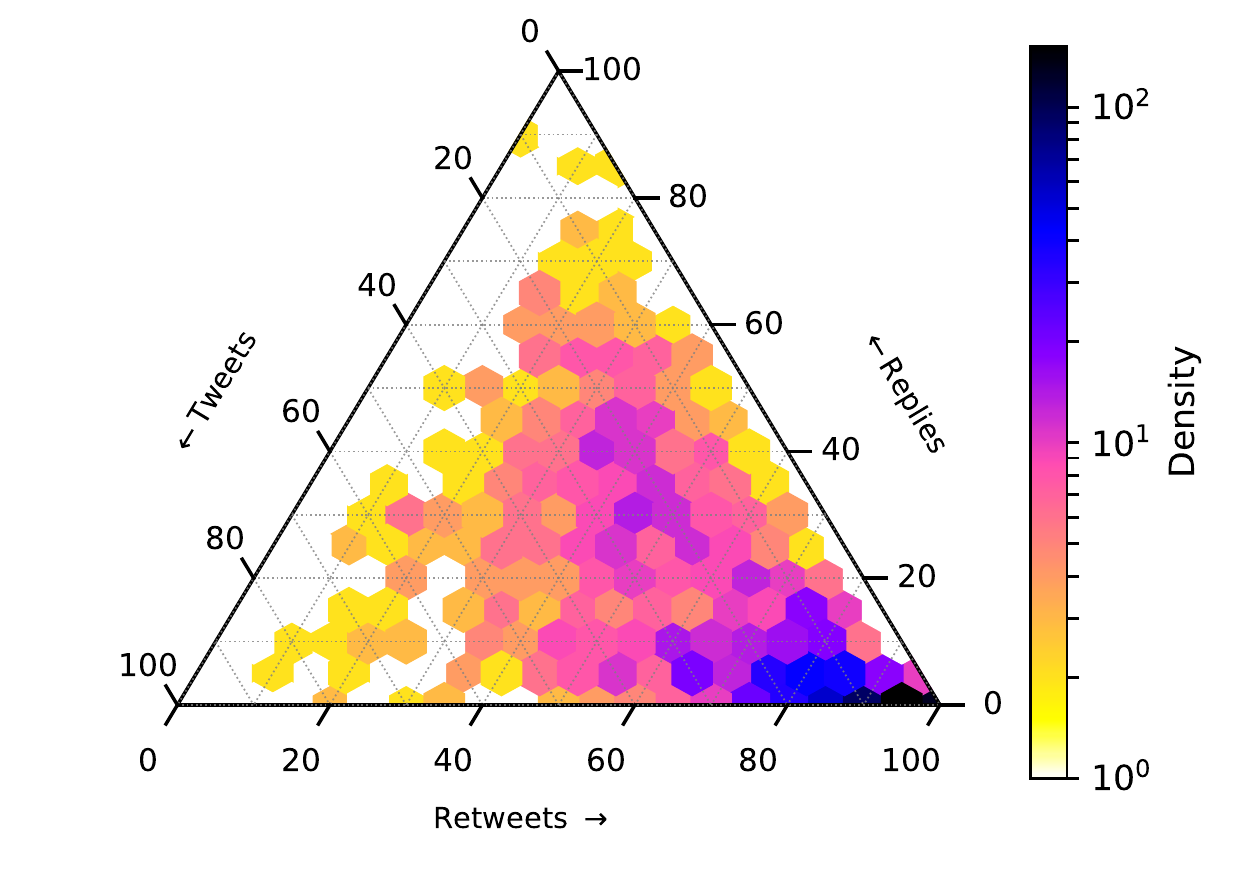} \\
{\sffamily\Large c}  \includegraphics[width=\halffig,valign=t]{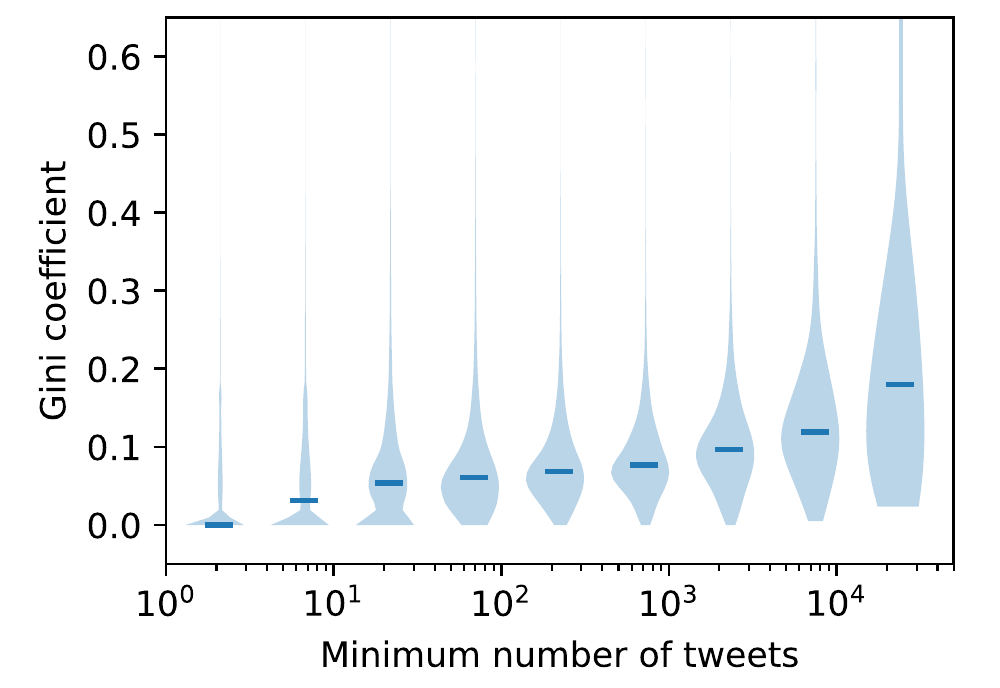} &
{\sffamily\Large d}  \includegraphics[width=\halffig,valign=t]{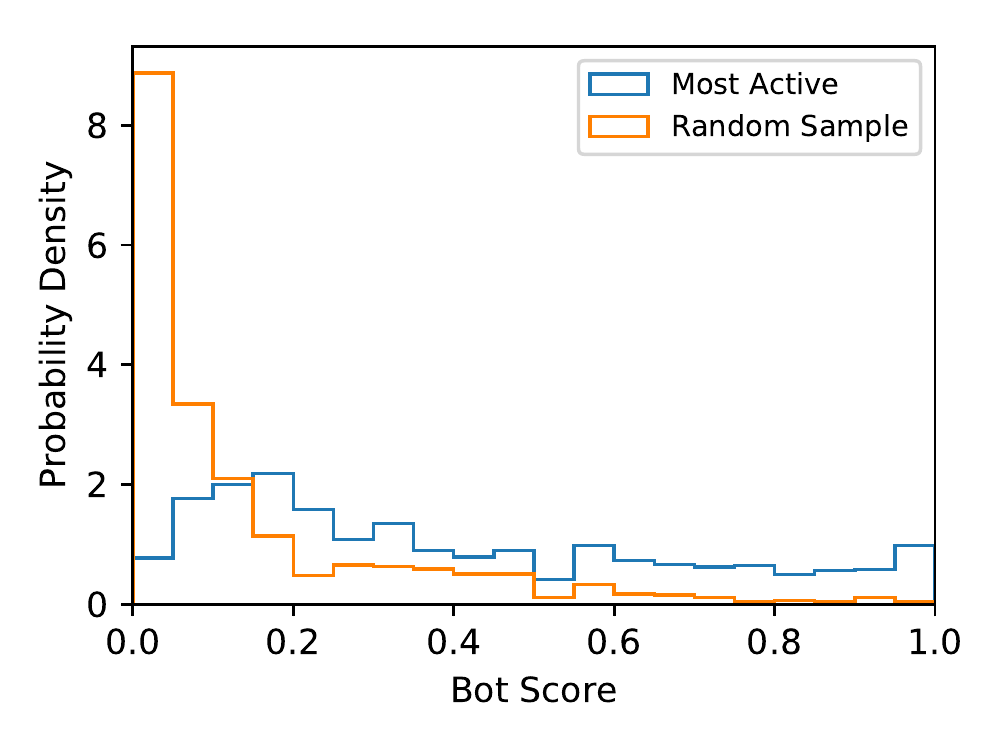} 
\end{tabular}
\caption{Anomalies. The distribution of types of tweet spreading articles from (a)~low-credibility and (b)~fact-checking sources are quite different. Each article is mapped along three axes  representing the percentages of different types of messages that share it: original tweets, retweets, and replies. When user Alice retweets a tweet by user Bob, the tweet is rebroadcast to all of Alice's followers, whereas when she replies to Bob's tweet, the reply is only seen by Bob. Color represents the number of articles in each bin, on a log-scale.
(c)~Correlation between popularity of articles from low-credibility sources and concentration of posting activity. We consider a collection of articles shared by a minimum number of tweets as a popularity group. For articles in each popularity group, a violin plot shows the distribution of Gini coefficients, which measure concentration of posts by few accounts (see Supplementary Materials). In violin plots, the width of a contour represents the probability of the corresponding value, and the median is marked by a colored line.
(d)~Bot score distributions for a random sample of 915 accounts who posted at least one link to a low-credibility source, and for the 961 ``super-spreaders'' that most actively shared content from low-credibility sources. The two groups have significantly different scores ($p<10^{-4}$ according to a Mann-Whitney $U$ test): super-spreaders are more likely bots.}
\label{fig:anomalies}
\end{figure}

Even though low-credibility and fact-checking sources show similar popularity distributions, we observe some anomalous patterns in the spread of low-credibility content. First, most articles by low-credibility sources spread through original tweets and retweets, while few are shared in replies (Fig.~\ref{fig:anomalies}(a)); this is different from articles by fact-checking sources, which are shared mainly via retweets but also replies (Fig.~\ref{fig:anomalies}(b)).
%
%We observe some anomalous patterns in the spread of low-credibility content. 
%First, although content from low-credibility and fact-checking sources has similar popularity distributions, this is the result of different diffusion mechanisms. Most low-credibility articles spread through original tweets and retweets, while few are shared in replies (Fig.~\ref{fig:anomalies}(a)); this is different from fact-checking articles, which are shared mainly via retweets but also replies (Fig.~\ref{fig:anomalies}(b)). 
In other words, the spreading patterns of low-credibility content are less ``conversational.''
Second, the more a story was tweeted, the more the tweets were concentrated in the hands of few accounts, who act as ``\emph{super-spreaders}'' (Fig.~\ref{fig:anomalies}(c)). This goes against the intuition that, as a story goes ``viral,'' and thus reaches a broader audience, the contribution of any individual account or group of accounts should matter less.
%
%Second, for the most popular articles from low-credibility sources, much of the spreading activity is concentrated around a small portion of accounts (Fig.~\ref{fig:anomalies}(c)), even though one would expect organic spread among many human users for viral articles. 
In fact, a single account can post the same low-credibility article hundreds or even thousands of times (see Supplementary Materials). This could suggest that the spread is amplified through automated means.

We hypothesize that the ``super-spreaders'' of low-credibility content are social bots, which are automatically posting links to articles, retweeting other accounts, or performing more sophisticated autonomous tasks, like following and replying to other users. To test this hypothesis, we used Botometer to evaluate the Twitter accounts that posted links to articles from low-credibility sources. For each account we computed a bot score (a number in the unit interval), which can be interpreted as the level of automation of that account. We used a threshold of 0.5 to classify an account as bot or human. 
%classified an accounts as likely bot or human by comparing its bot score to a threshold of 0.5. 
Details about the Botometer system and the threshold can be found in Methods. 
We first considered a random sample of the general population of accounts that shared at least one link to a low-credibility article. Only 6\% of accounts in the sample are labeled as bots using this method, but they are responsible for spreading 31\% of all tweets linking to low-credibility content, and 34\% of all articles from low-credibility sources. We then compared this group with a sample of the top most active accounts (``super-spreaders''), 33\% of which have been labeled as bot --- over five times as many (details in Supplementary Materials). 
Fig.~\ref{fig:anomalies}(d) 
%presents a further comparison between the two groups, 
confirms that the super-spreaders are significantly more likely to be bots compared to the general population of accounts who share low-credibility content. 
Because these results are based on a classification model, it could be possible that what we see in Fig.~\ref{fig:anomalies}(d) is due to bias in the way Botometer has been trained. For example, the model could learn to assign higher scores to more active accounts. We rule out this competing explanation by showing that higher bot scores cannot be attributed to this kind of bias in the learning model (see Supplementary Materials).

\begin{figure}
\centering
{\sffamily\Large a}  \includegraphics[width=\smallfig,valign=t]{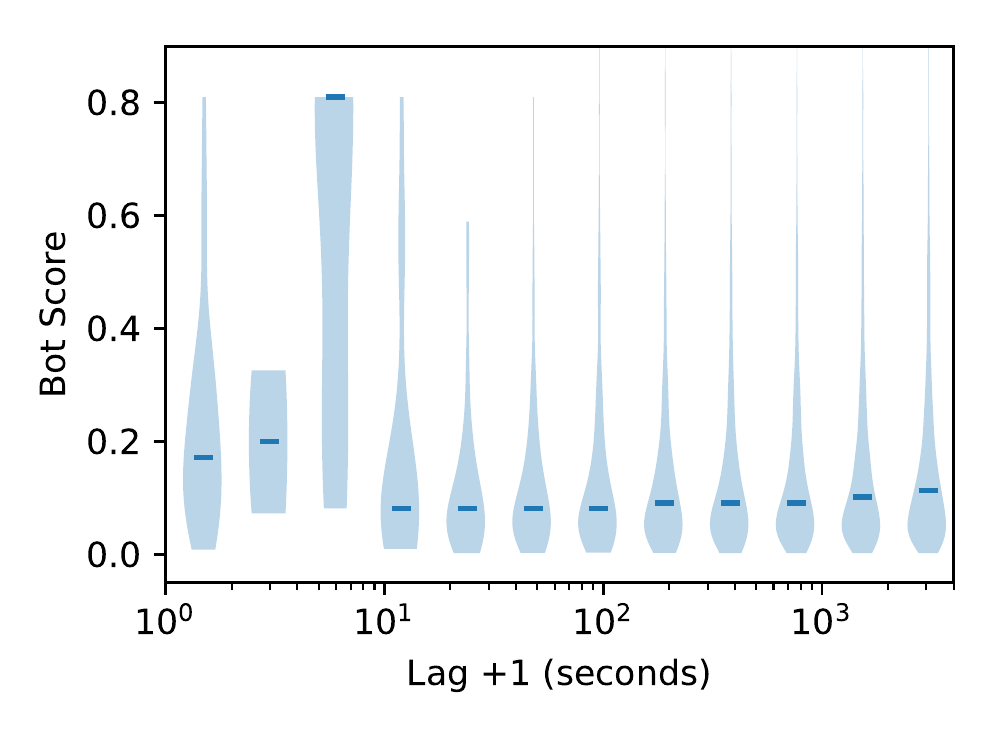} \\
{\sffamily\Large b} \includegraphics[width=\smallfig,valign=t]{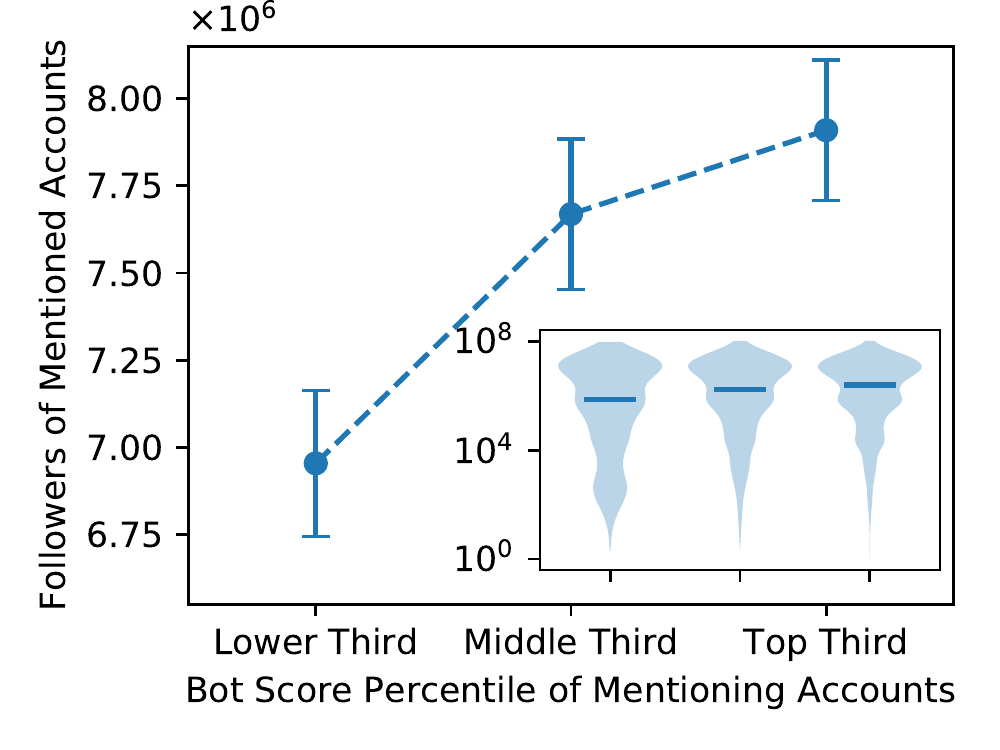}
\caption{Bot strategies. 
(a)~Early bot support after a viral low-credibility article is first shared. We consider a sample of 60,000 accounts that participate in the spread of the 1,000 most viral stories from low-credibility sources. We align the times when each article first appears. We focus on a one-hour early spreading phase following each of these events, and divide it into logarithmic lag intervals. The plot shows the bot score distribution for accounts sharing the articles during each of these lag intervals. 
(b)~Targeting of influentials. We plot the average number of followers of Twitter users who are mentioned (or replied to) by accounts that link to the most viral 1000 stories. The mentioning accounts are aggregated into three groups by bot score percentile. Error bars indicate standard errors. Inset: Distributions of follower counts for users mentioned by accounts in each percentile group.
}
\label{fig:bot-srategies}
\end{figure}

Given these anomalies, we submit that bots may play a critical role in driving the viral spread of content from low-credibility sources. To test this question, we examined whether bots tend to get involved at particular times in the spread of popular articles.
%We hypothesize that these bots play a critical role in driving the viral spread of low-credibility content. To test this conjecture,  we examined the different spreading phases of viral articles. In each of these phases we examined the accounts posting these articles. 
As shown in Fig.~\ref{fig:bot-srategies}(a), bots 
are prevalent in the first few seconds after an article is first published on Twitter. We conjecture that this 
early intervention exposes many users to low-credibility articles, 
increasing the chances than an article goes ``viral.''

We find that another strategy often used by bots is to mention influential users in tweets that link to low-credibility content. Bots seem to employ this targeting strategy repetitively; for example, a single account mentioned \texttt{@realDonaldTrump} in 19 tweets, each linking to the same false claim about millions of votes by illegal immigrants (see details in Supplementary Materials). For a systematic investigation, let us consider all tweets that mention or reply to a user and include a link to a viral article from a low-credibility source in our corpus. The number of followers is often used as a proxy for the influence of a Twitter user. 
As shown in Fig.~\ref{fig:bot-srategies}(b), in general tweets tend to mention popular people. However, accounts with the largest bot scores tend to mention users with a larger number of followers (median and average). 
A possible explanation for this strategy is that bots (or rather, their operators) target influential users with content from low-credibility sources, creating the appearance that it is widely shared. The hope is that these targets will then reshare the content to their followers, thus boosting its credibility.  

% GEO ANALYSIS MOVED TO SUPPLEMENTARY MATERIAL
We examined whether bot operators tended to target people in certain states by creating the appearance of users posting articles from those locations. We did find that the location patterns produced by bots are inconsistent with the geographic distribution of conversations on Twitter. However, we did not find evidence that bots sharing articles from low-credibility sources used this particular strategy to target swing states. Details of this analysis can be found in Supplementary Materials. 

\begin{figure}
\centerline{\includegraphics[width=\bigfig]{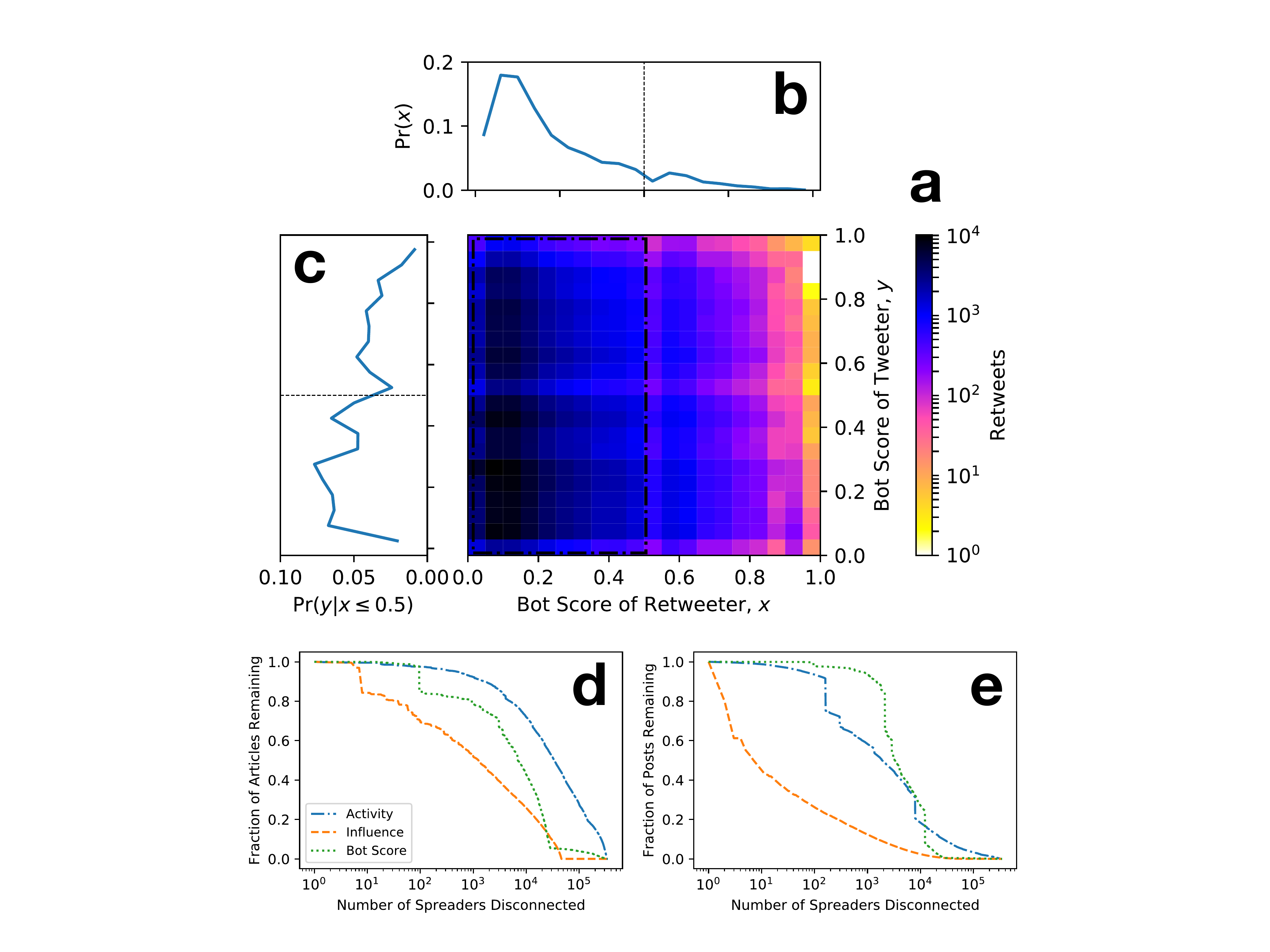}}
\caption{Impact of bots. 
(a)~Joint distribution of bot scores of accounts that retweeted links to low-credibility articles and accounts that had originally posted the links. Color represents the number of retweeted messages in each bin, on a log scale. (b)~The top projection shows the distributions of bot scores for retweeters, who are mostly human. (c)~The left projection shows the distributions of bot scores for accounts retweeted by likely humans (score below 0.5), with a significant portion of likely bots.
(d,e)~Dismantling the low-credibility content diffusion network. The priority of disconnected nodes is determined by ranking accounts on the basis of different characteristics. 
The remaining fraction of (d)~unique articles from low-credibility sources and (e)~retweets linking to those articles is plotted versus the number of disconnected nodes.
}
\label{fig:impact}
\end{figure}

Having found that bots are employed in ways that appear to drive the viral spread of low-credibility articles, let us explore how humans interact with the content shared by bots, which may provide insight into whether and how bots are able to affect public opinion. Fig.~\ref{fig:impact}(a) shows who retweets whom: humans do most of the retweeting (Fig.~\ref{fig:impact}(b)), and they retweet articles posted by bots as much as by other humans (Fig.~\ref{fig:impact}(c)). This suggests that collectively, people are not able to discriminate between low-credibility content shared by humans versus social bots. It also means that when we observe many accounts exposed to low-credibility information, these are not just bots (re)tweeting it. In fact, we find that the volume of tweets by likely humans scales super-linearly with the volume by likely bots, suggesting that the reach of these articles among humans is \emph{amplified} by social bots. In other words, each amount of sharing activity by bots tends to trigger a disproportionate amount of human engagement. The same amplification effect is not observed for articles from fact-checking sources. Details are presented in Supplementary Materials. 

Another way to assess the impact of bots in the spread of low-credibility content is to examine their critical role within the diffusion network. Let us focus on the retweet network~\cite{Shao2018anatomy}, where nodes are accounts and connections represents retweets of messages with links to stories --- just like the networks in Fig.~\ref{fig:a-pdf-tweets-and-users}(b,c), but aggregating across all articles from low-credibility sources. We apply a network dismantling procedure~\cite{albert2000error}: we disconnect one node at a time and analyze the resulting decrease in the total volume of retweets and in the total number of unique articles. The more these quantities are reduced by disconnecting a small number of nodes, the more critical those nodes are in the network.
We prioritize accounts to disconnect based on bot score and, for comparison, also based on retweeting activity and influence. Further details can be found in Methods. Unsurprisingly, Fig.~\ref{fig:impact}(d,e) shows that influential nodes are most critical. The most influential nodes are unlikely to be bots. Disconnecting nodes with high bot score is the second-best strategy for reducing low-credibility articles (Fig.~\ref{fig:impact}(d)). For reducing overall post volume, this strategy performs well when more than 10\% of nodes are disconnected (Fig.~\ref{fig:impact}(e)). Disconnecting active nodes is not as efficient a strategy for reducing low-credibility articles. 
These results show that bots are critical in the diffusion network: disconnecting likely bots results in the removal of most low-credibility content. 

\begin{figure}
\centerline{\includegraphics[width=\textwidth]{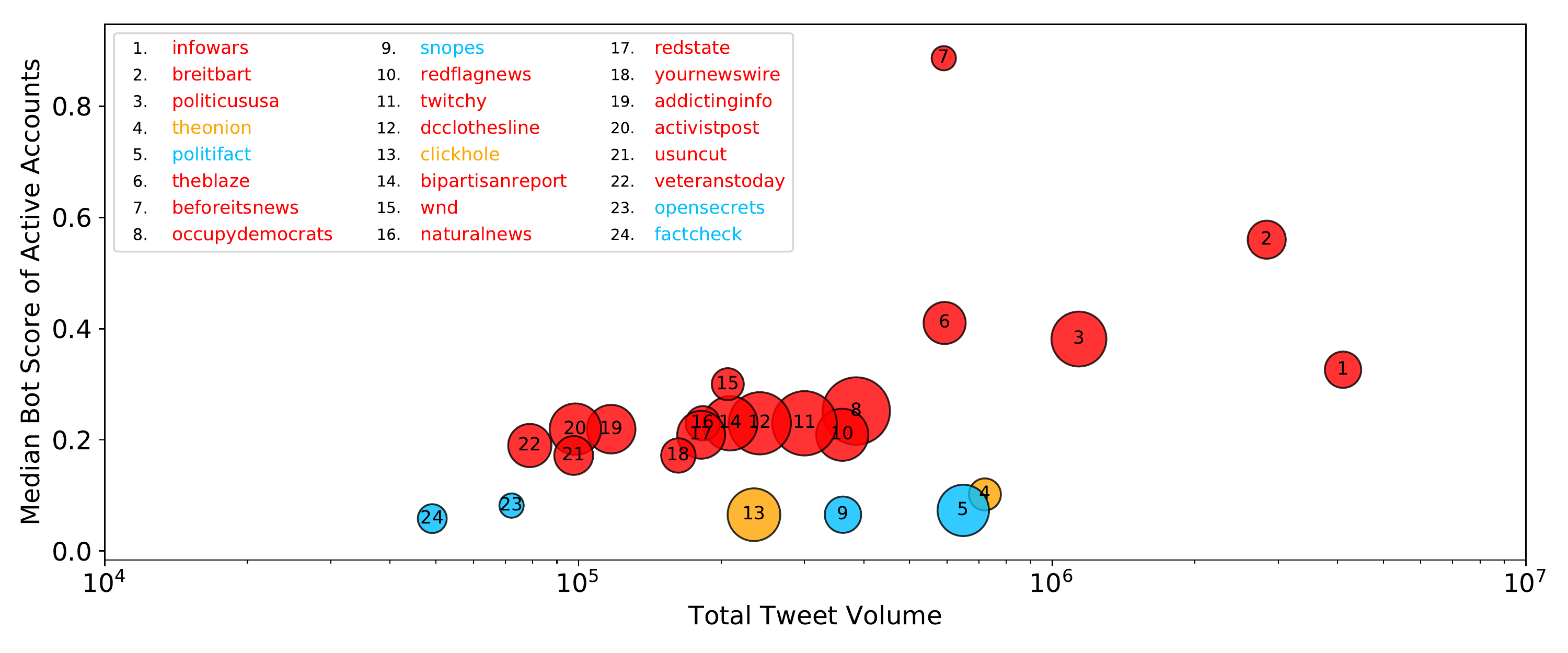}}
\caption{Popularity and bot support for the top sources. Satire websites are shown in orange,  fact-checking sites in blue, and low-credibility sources in red. Popularity is measured by total tweet volume (horizontal axis) and median number of tweets per article (circle area). Bot support is gauged by the median bot score of the 100 most active accounts posting links to articles from each source (vertical axis). Low-credibility sources have greater support by bots, as well as greater median and/or total volume in many cases. 
}
\label{fig:bots-in-sites}
\end{figure}

Finally, we compared the extent to which social bots 
% successfully manipulate the information ecosystem in 
disseminate content from different low-credibility sources. We considered the most popular sources in terms of median and aggregate article posts, and measured the bot scores of the accounts that most actively spread their content. As shown in Fig.~\ref{fig:bots-in-sites}, one site (\url{beforeitsnews.com}) stands out for the high degree of automation, 
% in terms of manipulation, 
but other popular low-credibility sources also have many bots among their promoters. The dissemination of content from satire sites like \textit{The Onion} and fact-checking websites does not display the same level of automation. 

\mysection{Discussion}

Our analysis provides quantitative empirical evidence of the key role played by social bots in the spread of low-credibility content. Relatively few accounts are responsible for a large share of the traffic that carries misinformation. These accounts are likely bots, and we uncovered two manipulation strategies they use. First, bots are particularly active in amplifying content in the very early spreading moments, before an article goes ``viral.'' Second, bots target influential users through replies and mentions. 
%Finally, bots may disguise their geographic locations. 
People are vulnerable to these kinds of manipulation, retweeting bots who post low-credibility content just as much as they retweet other humans. As a result, bots amplify the reach of low-credibility content, to the point that it is statistically indistinguishable from that of fact-checking articles. Successful low-credibility sources in the U.S., including those on both ends of the political spectrum, are heavily supported by social bots. Social media platforms are beginning to acknowledge these problems and deploy countermeasures, although their effectiveness is hard to evaluate~\cite{FB-INFO-OPS,FB-LOW-Q,fake-news-manifesto}.

The present findings are not in contradiction with the recent work by Vosoughi \textit{et al.}~\cite{Vosoughi1146}. Their analysis is based on a small subset of articles that are fact-checked, whereas the present work considers a much broader set of articles from low-credibility sources, most of which are not fact-checked. In addition, the analysis of Vosoughi \textit{et al.} did not consider an important mechanism by which bots can amplify the spread of an article, namely, by resharing links originally posted by human accounts. Because of these two methodological differences, the present analysis provides new evidence about the role played by bots.

Our results are \emph{robust} with respect to various choices. 
First, using a more restrictive criterion for selecting low-credibility sources, based on a consensus among several news and fact-checking organizations (see Methods), yields qualitatively similar results, leading to the same conclusions.
Second, our analysis about active spreaders of low-credibility content being likely bots is robust with respect to the activity threshold used to identify the most active spreaders. Furthermore, activity and bot scores are uncorrelated with account activity volume. 
Third, the conclusions are not affected by the use of different bot-score thresholds to separate social bots and human accounts.
Details about all of these robustness analyses can be found in Supplementary Materials. 

Our findings demonstrate that social bots are an effective tool to manipulate social media. While the present study focuses on the spread of low-credibility content, such as false news, conspiracy theories, and junk science, similar bot strategies may be used to spread other types of malicious content, such as malware. And although our spreading data is collected from Twitter, there is no reason to believe that the same kind of abuse is not taking place on other digital platforms as well. In fact, viral conspiracy theories spread on Facebook~\cite{del2016spreading} among the followers of pages that, like social bots, can easily be managed automatically and anonymously. Furthermore, just like on Twitter, false claims on Facebook are as likely to go viral as reliable news~\cite{low_quality_nhb2017}. While the difficulty to access spreading data on platforms like Facebook is a concern, the growing popularity of ephemeral social media like Snapchat may make future studies of this type of abuse all but impossible.

The results presented here suggest that curbing social bots may be an effective strategy for mitigating the spread of low-credibility content, and that the bot score might provide a useful signal to prioritize accounts for further review. Progress in this direction may be accelerated through partnerships between social media platforms and academic research~\cite{fake-news-manifesto}.  For example, our lab and others are developing machine learning algorithms to detect social bots~\cite{socialbots-CACM, socialbots-IEEE-DARPA, botornot_icwsm17}. The deployment of such tools is fraught with peril, however. While platforms have the right to enforce their terms of service, which forbid impersonation and deception, algorithms do make mistakes. Even a single false-positive error leading to the suspension of a legitimate account may foster valid concerns about censorship. This justifies current human-in-the-loop solutions, which unfortunately do not scale with the volume of abuse that is enabled by software. It is therefore imperative to support research both on improved abuse detection algorithms and on countermeasures that take into account the complex interplay between the cognitive and technological factors that favor the spread of misinformation~\cite{LEWANDOWSKY2017353}. 

An alternative strategy would be to employ CAPTCHAs~\cite{vonAhn2003}, challenge-response tests to determine whether a user is human. CAPTCHAs have been deployed widely and successfully to combat email spam and other types of online abuse. Their use to limit automatic posting or resharing of news links could help stem bot abuse by increasing its cost, but also add undesirable friction to benign applications of automation by legitimate entities, such as news media and emergency response coordinators. These are hard trade-offs that must be studied carefully as we contemplate ways to address the fake news epidemics.

\mysection{Methods}

Data about articles shared on Twitter was collected through Hoaxy, an open platform developed at Indiana University to track the spread of claims and fact checking~\cite{Shao2018anatomy}. A search engine, interactive visualizations, and open-source software are freely available (\url{hoaxy.iuni.iu.edu}). The data is accessible through a public API. Further details are presented in Supplementary Materials.

We started the collection in mid-May 2016 with 70 low-credibility sites and added 50 more in mid-December 2016. The collection period for the present analysis extends until the end of March 2017. During this time, we collected 389,569 articles from these 120 sites. We also tracked 15,053 stories published by independent fact-checking organizations, such as \url{snopes.com}, \url{politifact.com}, and \url{factcheck.org}. 
We did not exclude satire because many low-credibility sources label their content as satirical, 
and % making the distinction problematic. Furthermore, 
satire that goes ``viral'' is sometimes mistaken for real news. \textit{The Onion} is the satirical source with the highest total volume of shares. We repeated our analyses of most viral articles (e.g., Fig.~\ref{fig:bot-srategies}(a)) with articles from \url{theonion.com} excluded and the results were not affected. 

The full list of 120 sources is reported in Supplementary Materials. We also repeated the analysis using a more restrictive criterion for selecting low-credibility sources, based on a consensus among three or more news and fact-checking organizations. This yields 327,840 articles (86\% of the total) from 65 low-credibility sources, also listed in Supplementary Materials, where we show that the results are robust with respect to these different source selection criteria.  

Our analysis does not require a complete list of low-credibility sources, but does rely on the assumption that many articles published by these sources can be classified as some kind of misinformation or unsubstantiated information. To validate this assumption, we checked the content of a random sample of articles.  For the purpose of this verification, we adopted a definition of ``misinformation'' that follows industry convention and includes the following classes: fabricated content, manipulated content, imposter content, false context, misleading content, false connection, and satire~\cite{Wardle_FirstDraft_2017}. To these seven categories we also added articles whose claims could not be verified. We found that fewer that 15\% of articles could be verified. More details are available in Supplementary Materials.

Using the filtering endpoint of Twitter's public streaming API, we collected 13,617,425 public posts that included links to articles from low-credibility sources and 1,133,674 public posts linking to fact checks. This is the \emph{complete} set of tweets linking to these articles in the study period, and not a sample (see Supplementary Materials for details). We extracted metadata about the source of each link, the account that shared it, the original poster in case of retweet or quoted tweet, and any users mentioned or replied to in the tweet.

We transformed links to canonical URLs to merge different links referring to the same article. This happens mainly due to shortening services (44\% links are redirected) and extra parameters (34\% of URLs contain analytics tracking parameters), but we also found websites that use duplicate domains and snapshot services. Canonical URLs were obtained by resolving redirection and removing analytics parameters.

The bot score of Twitter accounts is computed using the Botometer classifier, which evaluates the extent to which an account exhibits similarity to the characteristics of social bots~\cite{botornot_icwsm17}. The system is based on a supervised machine learning algorithm leveraging more than a thousand features extracted from public data and meta-data about Twitter accounts. These features include various descriptors of information diffusion networks, user metadata, friend statistics, temporal patterns of activity, part-of-speech and sentiment analysis. The classifier is trained using publicly available datasets of tens of thousands of Twitter users that include both humans and bots of varying sophistication. The Botometer system is available through a public API (\url{botometer.iuni.iu.edu}). It has also been employed in other studies~\cite{Vosoughi1146,Pew2018} and is widely adopted, serving hundreds of thousand requests daily. 

For the present analysis, we use the Twitter Search API to collect up to 200 of an account's most recent tweets and up to 100 of the most recent tweets mentioning the account. From this data we extract the features used by the Botometer classifier. We use logistic calibration to make the bot scores calculated by the classifier easier to interpret as confidence levels (see Supplementary Materials).

There is no crisp, binary classification of accounts as human or bot, as there are many types of bots and humans using different levels of automation. Accordingly, Botometer provides a score (percentage) rather than a binary classification. Nevertheless, the model can effectively discriminate between human and bot accounts of different nature; five-fold cross-validation yields an area under the ROC curve of 94\%~\cite{botornot_icwsm17}. A value of 50\% indicates random accuracy and 100\% means perfect accuracy. 
When a binary classification is needed, we use a threshold of 0.5, which maximizes accuracy~\cite{botornot_icwsm17}. See Supplementary Materials for further details about bot classification and its robustness. 

In the targeting analysis (Fig.~\ref{fig:bot-srategies}(b)), we exclude mentions of sources using the pattern ``via @screen\_name.''

The network studied in the dismantling analysis (Fig.~\ref{fig:impact}(d,e)) is based on retweets with links to articles from low-credibility sources, posted before the election (May 16--Nov 7, 2016). The network has 630,368 nodes (accounts) and 2,236,041 directed edges. Each edge is weighted by the number of retweets between the same pair of accounts. When an account is disconnected, all of its incoming and outgoing edges are removed. When we disconnect a retweeting node $i$ that was in turn retweeted by some node $j$, only $i$ is removed because in the Twitter metadata, each retweet connects directly to the account that originated the tweet. Given the directionality of edges, retweeting activity is measured by node in-strength (weighted in-degree) and influence by out-strength (weighted out-degree).  

\paragraph{Acknowledgments.} 
We are grateful to Ben Serrette and Valentin Pentchev of the Indiana University Network Science Institute (\url{iuni.iu.edu}), as well as Lei Wang for supporting the development of the Hoaxy platform. Pik-Mai Hui assisted with the dismantling analysis. Clayton A. Davis developed the Botometer API. Nic Dias provided assistance with claim verification. We are also indebted to Twitter for providing data through their API. C.S. thanks the Center for Complex Networks and Systems Research (\url{cnets.indiana.edu}) for the hospitality during his visit at the Indiana University School of Informatics, Computing, and Engineering. He was supported by the China Scholarship Council. G.L.C. was supported by IUNI. The development of the Botometer platform was supported in part by DARPA (grant W911NF-12-1-0037). The development of the Hoaxy platform was supported in part by the Democracy Fund. A.F. and F.M. were supported in part by the James S. McDonnell Foundation (grant 220020274) and the National Science Foundation (award  CCF-1101743). The funders had no role in study design, data collection and analysis, decision to publish or preparation of the manuscript.

\appendix

\section*{Appendix: Supplementary Background}

Tracking abuse of social media has been a topic of intense research in recent years. The analysis in the main text leverages Hoaxy, a system focused on tracking the spread of links to articles from low-credibility and fact-checking sources~\cite{shao2016hoaxy}. Here we give a brief overview of other systems designed to monitor the spread of misinformation on social media. This is related to the problems of mining and detecting misinformation and fake news, which are the subjects of recent surveys~\cite{wu2016mining,shu2017fake}. 

Beginning with the detection of simple instances of political abuse like \emph{astroturfing}~\cite{Ratkiewicz2011}, researchers noted the need for automated tools for monitoring social media streams and detecting manipulation or misinformation. Several such systems have been proposed in recent years, each with a particular focus or a different approach. The Truthy system~\cite{Ratkiewicz2011} relied on network analysis techniques to classify memes, such as hashtags and links. TraceMiner~\cite{Wu2018TraceMiner} also models the propagation of messages, but by inferring embeddings of social media users with social network structures. The TweetCred system~\cite{Castillo2011,gupta2014tweetcred} focuses on content-based features and other kind of metadata, and distills a measure of overall information credibility. The Hierarchical Credibility Network~\cite{jin2014news} considers credibility as propagating through a three-layer network consisting of event, sub-events, and messages classified based on their features. 

Specific systems have been proposed to detect rumors~\cite{zubiaga2018detection}. These include RumorLens~\cite{Resnick2014},  TwitterTrails~\cite{metaxas2015using}, FactWatcher~\cite{Hassan2014}, and News Tracer~\cite{liu2015real}. The news verification capabilities of these systems range from completely automatic (TweetCred), to semi-automatic (TwitterTrails, RumorLens, News Tracer). In addition, some of them let the user explore the propagation of a rumor with an interactive dashboard (TwitterTrails, RumorLens). These systems vary in their capability to monitor the social media stream automatically, but in all cases the user is required to enter a seed rumor or keyword to operate them.

Our analysis is based on the spread of content from low-credibility sources rather than focusing on individual stories that are labeled as misinformation. Due to the impossibility to fact-check millions of articles, this approach of using sources as proxies for misinformation labels is increasingly adopted in the literature cited in the main text, and more~\cite{doi:10.1177/1461444817712086,starbird2017examining,Zannettou:2017:WCU:3131365.3131390,10.1371/journal.pone.0118093,guess2018selective}. 

Since misinformation can be propagated by coordinated online campaigns, it is important to detect whether a meme is being artificially promoted. Machine learning has been applied successfully to the task of early discriminating between trending memes that are either organic or promoted by means of advertisement~\cite{Varol2017epjds}.

Finally, there is a growing body of research on social bot detection. The level of sophistication of bot-based manipulation can vary greatly~\cite{boshmaf2011socialbot}. As discussed in the main text, there is a large gray area between human and completely automated accounts. So-called cyborgs are accounts used to amplify content generated by humans~\cite{chu2010tweeting}. It is possible that a significant portion of the manipulation discussed in this paper, aimed to amplify the spread of low-credibility content, is carried out by this kind of bot. The Botometer system used in this paper has been publicly available for a few years~\cite{Davis16BotOrNot}. Its earliest version was trained on simple spam bots, detected through a social honeypot system~\cite{webb2008social,lee2010uncovering}. The version used here was trained on public datasets that also included more sophisticated bots.

\section*{Appendix: Supplementary Methods}

\subsection*{List of Sources}

Our list of low-credibility sources was obtained by merging several lists compiled by third-party news and fact-checking organizations or experts. It should be noted that these lists were compiled independently of each other, and as a result they have uneven coverage. However, there is overlap between them. The full list of sources is shown in Table~\ref{tab:sources}. Some lists annotate sources in different categories. In the case of OpenSources (\url{www.opensources.co}), we only considered sources tagged with any of the following labels: \textit{fake, satire, bias, conspiracy, rumor, state, junksci, clickbait, hate.} In the case of Starbird's list of alternative domains~\cite{starbird2017examining}, we considered those with primary orientation coded as one of \textit{conspiracy theorists, political agenda, tabloid|clickbait news.} 

For robustness analysis (below), we also consider a ``consensus'' subset of sites that are each listed among low-credibility sources by at least three organizations or experts. This subset includes 65 sources, also shown in Table~\ref{tab:sources}. We track 10,663,818 tweets (79\% of the total) with links to 327,840 articles (86\% of the total) from consensus low-credibility source, generated by 1,135,167 accounts (84\% of the total). 

We additionally tracked the websites of seven independent fact-checking organizations: \url{badsatiretoday.com}, \url{factcheck.org}, \url{hoax-slayer.com},\footnote{\url{hoax-slayer.com} includes its older version \url{hoax-slayer.net}.}   \url{opensecrets.org}, \url{politifact.com}, \url{snopes.com}, and \url{truthorfiction.com}. In April 2017 we added \url{climatefeedback.org}, which does not affect the present analysis. 

% A new command to generate table header row for the following long table (sources)
\newcommand{\tableheadrow}{
    \bfseries Source &
    % Listed by Fake News Watch
    \bfseries \href{http://archive.is/4x42V}{FNW} &
    % Melissa Zimdars/OpenSources
    \bfseries \href{http://www.opensources.co/}{OS} &
    % The Daily Dot
    \bfseries \href{http://www.dailydot.com/layer8/fake-news-sites-list-facebook/}{DD} &
    % Listed by U.S. News
    \bfseries \href{http://www.usnews.com/news/national-news/articles/2016-11-14/avoid-these-fake-news-sites-at-all-costs}{US} &
    % Listed by New Republic
    \bfseries \href{https://newrepublic.com/article/118013/satire-news-websites-are-cashing-gullible-outraged-readers}{NR} &
    % Listed by CBS News
    \bfseries \href{http://www.cbsnews.com/pictures/dont-get-fooled-by-these-fake-news-sites/}{CBS} &
    % Listed by about.com
    \bfseries \href{http://urbanlegends.about.com/od/Fake-News/tp/A-Guide-to-Fake-News-Websites.htm}{UL} &
    % Listed by NPR
    \bfseries \href{http://www.npr.org/sections/alltechconsidered/2016/11/23/503146770/npr-finds-the-head-of-a-covert-fake-news-operation-in-the-suburbs}{NPR} &
    % Listed by Snopes Field Guide
    \bfseries \href{http://www.snopes.com/2016/01/14/fake-news-sites/}{Sn} &
    % Listed by Kate Starbird
    \bfseries \href{https://docs.google.com/spreadsheets/d/1lk3pFSc5wo3OfJc8ekONqO3MJCCigqe8SBSYwLYlHLo}{KS} &
    % Listed by Craig Silverman/BuzzFeed
    \bfseries \href{https://github.com/BuzzFeedNews/2017-12-fake-news-top-50/tree/master/data}{BF} &
    % Listed by Politifact
    \bfseries \href{http://www.politifact.com/punditfact/article/2017/apr/20/politifacts-guide-fake-news-websites-and-what-they/}{PF} &
    \bfseries Date \\
}

% 11 columns
% use a smaller font size for table contents
\begin{ThreePartTable}
\begin{footnotesize}
\begin{TableNotes}
\item[a] \url{newslo.com} and \url{politicops.com} are mirrors of \url{politicot.com}.
  \item[b] \url{thedcgazette.com} is a mirror of \url{dcgazette.com}. 
\end{TableNotes}
%% Fil's attempt to center wide table and adjust caption
\setlength\LTleft{-0.4in}
\setlength\LTright{-0.4in}
\setlength\LTcapwidth{\linewidth}
\begin{longtable}{lccccccccccccc}
\caption{\normalsize{Low-credibility sources. For each source, we indicate which lists include it. The lists are: Fake News Watch (FNW), OpenSources (OS), Daily Dot (DD), US News \& World Report (US), New Republic (NR), CBS, Urban Legends (UL), NPR, Snopes Field Guide (Sn), Starbird Alternative Domains (KS), BuzzFeed News (BF), and PolitiFact (PF). Table headers link to the original lists. The date indicates when Hoaxy started following a source: June 29 or December 20, 2016. Consensus sources (in three or more lists) are shown in italics.}}
% headers of first page
\\ % needed to avoid complaints, weird
\hline
\tableheadrow
\hline
\endfirsthead
    
% header for other pages (except first page), including the continue line and head row
\multicolumn{14}{l}{{\bfseries \tablename\ \thetable{} -- continued from previous page}} \\
\hline
\tableheadrow
\hline
\endhead
    
% footers of pages (except last page)
\hline
\multicolumn{14}{r}{{Continued on next page}}\\
\endfoot

% footers of last
% Table notes.
\hline 
\insertTableNotes
\endlastfoot
%
% table content generation
% no head line, because we need to self define them.
\begin{filecontents*}{sources2.csv}
%,,OS,,,,,,,,KS,CS,PF,
{\it 21stcenturywire.com},\checkmark,\checkmark,\checkmark,,,,,,,\checkmark,,,Jun
{\it 70news.wordpress.com},,\checkmark,\checkmark,,,\checkmark,,,,,,,Dec
{\it abcnews.com.co},,\checkmark,\checkmark,,,\checkmark,,,,,\checkmark,\checkmark,Dec
{\it activistpost.com},\checkmark,\checkmark,\checkmark,\checkmark,,,,,,\checkmark,,,Jun
{\it addictinginfo.org},,\checkmark,\checkmark,,,,,,,\checkmark,,,Dec
{\it americannews.com},\checkmark,\checkmark,\checkmark,\checkmark,,,,,,,,\checkmark,Jun
americannewsx.com,,\checkmark,,,,,,,,,,,Dec
amplifyingglass.com,\checkmark,,,,,,,,,,,,Jun
anonews.co,,,\checkmark,,,,,,,,,,Dec
{\it beforeitsnews.com},\checkmark,\checkmark,,\checkmark,,,,,,\checkmark,,\checkmark,Jun
bigamericannews.com,\checkmark,\checkmark,,,,,,,,,,,Jun
bipartisanreport.com,,\checkmark,\checkmark,,,,,,,,,,Dec
bluenationreview.com,,\checkmark,\checkmark,,,,,,,,,,Dec
{\it breitbart.com},,\checkmark,\checkmark,,,,,,,\checkmark,,,Dec
{\it burrardstreetjournal.com},,\checkmark,,,,\checkmark,,,,,\checkmark,,Dec
{\it callthecops.net},,\checkmark,\checkmark,,,,\checkmark,,,,,,Dec
christiantimes.com,,,,,,\checkmark,,,,,,,Dec
{\it christwire.org},\checkmark,\checkmark,\checkmark,,,,,,,,,,Jun
chronicle.su,\checkmark,\checkmark,,,,,,,,,,,Jun
{\it civictribune.com},\checkmark,\checkmark,\checkmark,,,\checkmark,,,,,\checkmark,\checkmark,Jun
{\it clickhole.com},\checkmark,\checkmark,\checkmark,\checkmark,,,,,,,,,Jun
coasttocoastam.com,\checkmark,,\checkmark,,,,,,,,,,Jun
collective-evolution.com,,,\checkmark,,,,,,,,,,Dec
{\it consciouslifenews.com},\checkmark,\checkmark,\checkmark,,,,,,,\checkmark,,,Jun
{\it conservativeoutfitters.com},\checkmark,\checkmark,\checkmark,,,,,,,,,,Dec
{\it countdowntozerotime.com},\checkmark,\checkmark,\checkmark,,,,,,,,,,Jun
counterpsyops.com,\checkmark,\checkmark,,,,,,,,,,,Jun
{\it creambmp.com},\checkmark,\checkmark,\checkmark,,,,,,,,,,Jun
{\it dailybuzzlive.com},\checkmark,\checkmark,,\checkmark,,,,,,,,\checkmark,Jun
{\it dailycurrant.com},\checkmark,\checkmark,,,,,\checkmark,,,,\checkmark,,Jun
dailynewsbin.com,,\checkmark,,,,,,,,,,,Dec
{\it dcclothesline.com},\checkmark,\checkmark,,,,,,,,\checkmark,,,Jun
demyx.com,,,,,\checkmark,,,,,,,,Dec
{\it denverguardian.com},,\checkmark,,,,,,\checkmark,,,\checkmark,,Dec
derfmagazine.com,\checkmark,\checkmark,,,,,,,,,,,Jun
{\it disclose.tv},\checkmark,\checkmark,,\checkmark,,,,,,,,\checkmark,Jun
{\it duffelblog.com},\checkmark,,\checkmark,\checkmark,,,,,,,,\checkmark,Jun
duhprogressive.com,\checkmark,\checkmark,,,,,,,,,,,Jun
{\it empireherald.com},,\checkmark,,,,\checkmark,,,,,\checkmark,\checkmark,Dec
{\it empirenews.net},\checkmark,\checkmark,\checkmark,,,\checkmark,\checkmark,,\checkmark,,\checkmark,\checkmark,Jun
{\it empiresports.co},\checkmark,\checkmark,,,\checkmark,,\checkmark,,\checkmark,,\checkmark,\checkmark,Jun
{\it en.mediamass.net},\checkmark,,\checkmark,,\checkmark,,\checkmark,,,,,,Jun
endingthefed.com,,\checkmark,,,,,,,,,,,Dec
{\it enduringvision.com},\checkmark,\checkmark,\checkmark,,,,,,,,,,Jun
flyheight.com,,\checkmark,,,,,,,,,,,Dec
fprnradio.com,\checkmark,\checkmark,,,,,,,,,,,Jun
{\it freewoodpost.com},,\checkmark,,,,,\checkmark,,,,\checkmark,\checkmark,Dec
geoengineeringwatch.org,\checkmark,\checkmark,,,,,,,,,,,Jun
{\it globalassociatednews.com},,,,,\checkmark,,\checkmark,,,,\checkmark,,Dec
{\it globalresearch.ca},\checkmark,\checkmark,,,,,,,,\checkmark,,,Jun
gomerblog.com,\checkmark,,,,,,,,,,,,Jun
{\it govtslaves.info},\checkmark,\checkmark,,,,,,,,\checkmark,,,Jun
gulagbound.com,\checkmark,\checkmark,,,,,,,,,,,Jun
hangthebankers.com,\checkmark,\checkmark,,,,,,,,,,,Jun
humansarefree.com,\checkmark,\checkmark,,,,,,,,,,,Jun
{\it huzlers.com},\checkmark,\checkmark,,,\checkmark,\checkmark,\checkmark,,\checkmark,,\checkmark,\checkmark,Jun
ifyouonlynews.com,,\checkmark,,,,\checkmark,,,,,,,Dec
{\it infowars.com},\checkmark,\checkmark,\checkmark,\checkmark,,\checkmark,,,,\checkmark,,,Jun
{\it intellihub.com},\checkmark,\checkmark,,,,,,,,\checkmark,,,Jun
itaglive.com,\checkmark,,,,,,,,,,,,Jun
jonesreport.com,\checkmark,\checkmark,,,,,,,,,,,Jun
{\it lewrockwell.com},\checkmark,\checkmark,,,,,,,,\checkmark,,,Jun
liberalamerica.org,,\checkmark,,,,,,,,,,,Dec
libertymovementradio.com,\checkmark,\checkmark,,,,,,,,,,,Jun
libertytalk.fm,\checkmark,\checkmark,,,,,,,,,,,Jun
libertyvideos.org,\checkmark,\checkmark,,,,,,,,,,,Jun
lightlybraisedturnip.com,,,,,\checkmark,,,,,,,,Dec
{\it nationalreport.net},\checkmark,\checkmark,\checkmark,,\checkmark,\checkmark,\checkmark,\checkmark,\checkmark,,\checkmark,\checkmark,Jun
{\it naturalnews.com},\checkmark,\checkmark,\checkmark,\checkmark,,,,,,,,,Jun
{\it ncscooper.com},,\checkmark,,,,,,,\checkmark,,\checkmark,,Dec
{\it newsbiscuit.com},\checkmark,\checkmark,\checkmark,,,,,,,,\checkmark,,Jun
%newsexaminer.net,,\checkmark,,,,,,,\checkmark,,\checkmark,\checkmark,Dec
{\it newslo.com}\tnote{a},\checkmark,\checkmark,\checkmark,\checkmark,,,,,,,\checkmark,\checkmark,Jun
{\it newsmutiny.com},\checkmark,\checkmark,\checkmark,,,,,,,,,,Jun
newswire-24.com,\checkmark,\checkmark,,,,,,,,,,,Jun
{\it nodisinfo.com},\checkmark,\checkmark,,,,,,,,\checkmark,,,Jun
{\it now8news.com},,\checkmark,,,,\checkmark,,,\checkmark,,\checkmark,\checkmark,Dec
nowtheendbegins.com,\checkmark,\checkmark,,,,,,,,,,,Jun
{\it occupydemocrats.com},,\checkmark,\checkmark,,,,,,,\checkmark,,,Dec
other98.com,,\checkmark,\checkmark,,,,,,,,,,Dec
pakalertpress.com,\checkmark,\checkmark,,,,,,,,,,,Jun
politicalblindspot.com,\checkmark,\checkmark,,,,,,,,,,,Jun
politicalears.com,\checkmark,\checkmark,,,,,,,,,,,Jun
{\it politicops.com}\tnote{a},\checkmark,\checkmark,,,,\checkmark,,,,,\checkmark,\checkmark,Jun
politicususa.com,,\checkmark,,,,,,,,,,,Dec
prisonplanet.com,\checkmark,\checkmark,,,,,,,,,,,Jun
{\it react365.com},,\checkmark,,,,\checkmark,,,\checkmark,,\checkmark,\checkmark,Dec
realfarmacy.com,\checkmark,\checkmark,,,,,,,,,,,Jun
{\it realnewsrightnow.com},\checkmark,\checkmark,\checkmark,,,\checkmark,,,,,\checkmark,\checkmark,Jun
{\it redflagnews.com},\checkmark,\checkmark,,\checkmark,,,,,,\checkmark,,,Jun
redstate.com,,\checkmark,\checkmark,,,,,,,,,,Dec
{\it rilenews.com},\checkmark,\checkmark,\checkmark,,,\checkmark,,,,,\checkmark,\checkmark,Jun
rockcitytimes.com,\checkmark,,,,,,,,,,,,Jun
{\it satiratribune.com},,\checkmark,,,,,,,\checkmark,,\checkmark,\checkmark,Dec
{\it stuppid.com},,\checkmark,,,,,,,\checkmark,,\checkmark,,Dec
theblaze.com,,\checkmark,,,,,,,,,,,Dec
{\it thebostontribune.com},,\checkmark,,,,\checkmark,,,,,\checkmark,,Dec
{\it thedailysheeple.com},\checkmark,\checkmark,,,,,,,,\checkmark,,,Jun
{\it thedcgazette.com}\tnote{b},\checkmark,,\checkmark,\checkmark,,\checkmark,,,,,,,Jun
{\it thefreethoughtproject.com},,\checkmark,\checkmark,,,,,,,\checkmark,,,Dec
thelapine.ca,\checkmark,,,,,,\checkmark,,,,,,Jun
{\it thenewsnerd.com},\checkmark,\checkmark,,,\checkmark,,,,,,\checkmark,,Jun
{\it theonion.com},\checkmark,\checkmark,\checkmark,\checkmark,\checkmark,\checkmark,\checkmark,,,,,,Jun
theracketreport.com,,\checkmark,,,,,\checkmark,,,,\checkmark,\checkmark,Dec
{\it therundownlive.com},\checkmark,\checkmark,,,,,,,,,,,Jun
{\it thespoof.com},\checkmark,\checkmark,,,,,\checkmark,,,,,,Jun
theuspatriot.com,\checkmark,\checkmark,,,,,,,,,,,Jun
truthfrequencyradio.com,\checkmark,\checkmark,,,,,,,,,,,Jun
twitchy.com,,,\checkmark,,,,,,,,,,Dec
unconfirmedsources.com,\checkmark,\checkmark,,,,,,,,,,,Jun
USAToday.com.co,,,,,,,,\checkmark,\checkmark,,,,Dec
usuncut.com,,\checkmark,\checkmark,,,,,,,,,,Dec
{\it veteranstoday.com},\checkmark,\checkmark,,,,,,,,\checkmark,,,Jun
wakingupwisconsin.com,\checkmark,\checkmark,,,,,,,,,,,Jun
{\it weeklyworldnews.com},\checkmark,\checkmark,,\checkmark,,,\checkmark,,,,,,Jun
wideawakeamerica.com,\checkmark,,,,,,,,,,,,Jun
winningdemocrats.com,,\checkmark,,,,,,,,,,,Dec
{\it witscience.org},\checkmark,\checkmark,,,,,,,,,\checkmark,,Jun
wnd.com,,\checkmark,,,,,,,,,,,Dec
{\it worldnewsdailyreport.com},\checkmark,\checkmark,\checkmark,,,,\checkmark,,\checkmark,,\checkmark,\checkmark,Jun
{\it worldtruth.tv},\checkmark,\checkmark,,\checkmark,,,,,,\checkmark,,\checkmark,Jun
{\it yournewswire.com},,\checkmark,,,,\checkmark,,,,\checkmark,\checkmark,\checkmark,Dec
\end{filecontents*}
\csvreader[no head]{sources2.csv}{}
% specify the columns to render
{
    \csvcoli &
    \csvcolii &
    \csvcoliii &
    \csvcoliv &
    \csvcolv &
    \csvcolvi &
    \csvcolvii &
    \csvcolviii &
    \csvcolix &
    \csvcolx &
    \csvcolxi &
    \csvcolxii &
    \csvcolxiii &
    \csvcolxiv \\
}
\label{tab:sources}
\end{longtable}
\end{footnotesize}
\end{ThreePartTable}

\subsection*{Hoaxy Data}

The back-end component of Hoaxy collects public tweets that link to a predefined list of websites. The use of the ``POST statuses/filter'' endpoint of the Twitter streaming API, together with the total volume of tweets collected, which is well below 1\% of all public tweets, guarantee that we obtain all tweets linking to the sites in our list, and not just a sample of the tweets with these links. In addition, Hoaxy crawls all tracked websites and indexes all their articles, supporting a full-text search engine that allows users to find articles matching a given query. Furthermore, users can select subsets of these articles to visualize their spread on Twitter. To this end, Hoaxy matches the indexed articles with the tweets in our database and constructs networks based on retweets, mentions, replies, and quoted tweets. The front-end visualizes these networks interactively, allowing users to explore the accounts (nodes) and the tweets (edges) that make up these networks. The system makes all the data accessible to the public through a website (\url{hoaxy.iuni.iu.edu}) and an API.

\begin{figure}
\centerline{\includegraphics[width=\bigfig]{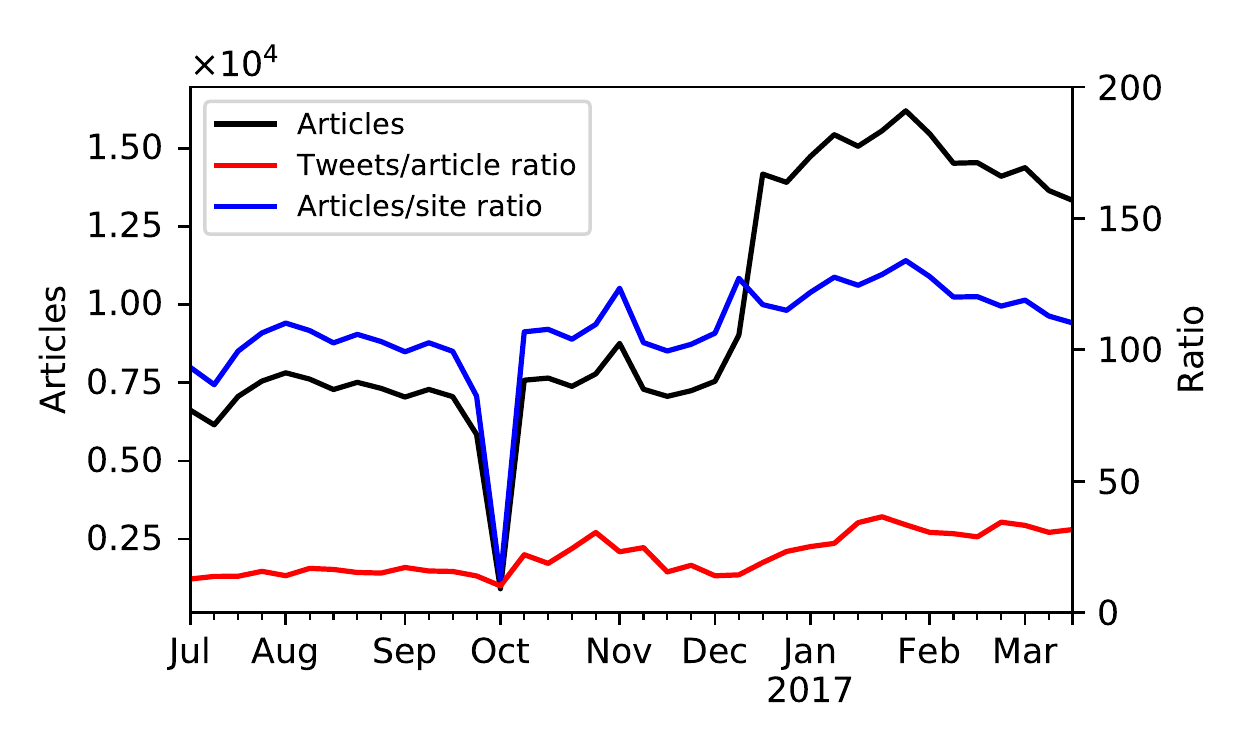}}
\caption{Weekly tweeted low-credibility articles, tweets/article ratio and articles/site ratio. The collection was briefly interrupted in October 2016. In December 2016 we expanded the set of low-credibility sources, from 70 to 120 websites.}
\label{fig:a-timeline}
\end{figure}

\begin{figure}
\centerline{\includegraphics[width=\bigfig]{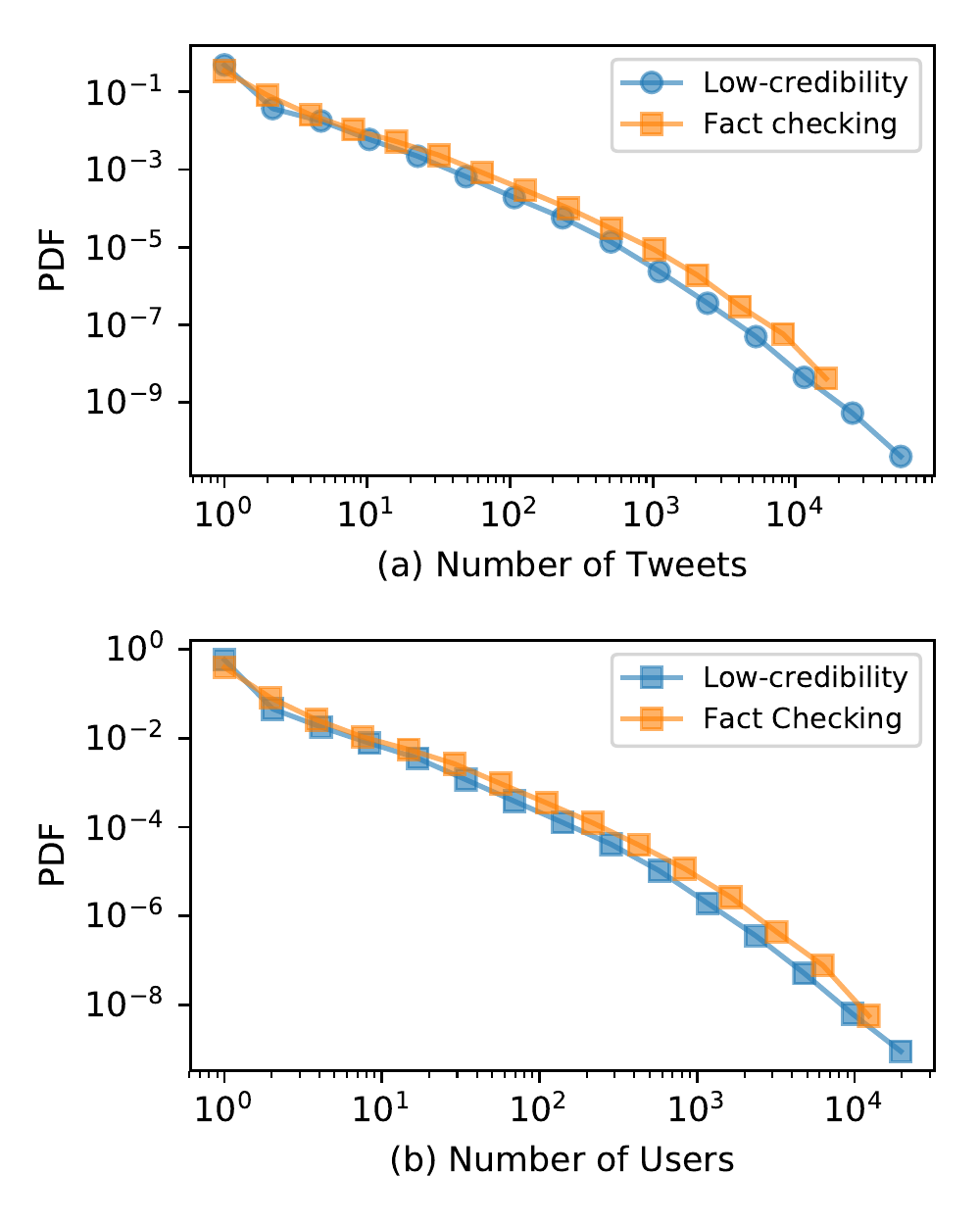}}
\caption{Probability distributions of popularity of articles from low-credibility and fact-checking sources, measured by (a)~the number of tweets and (b)~the number of accounts sharing links to an article.}
\label{fig:pop-distr}
\end{figure}

Our analysis focuses on the period from mid-May 2016 to the end of March 2017. During this time, we collected 15,053 and 389,569 articles from fact-checking and low-credibility sources, respectively. The Hoaxy system collected 1,133,674 public posts that included links to fact checks and 13,617,425 public posts linking to low-credibility articles. As shown in Fig.~\ref{fig:a-timeline}, low-credibility websites each produced approximately 100 articles per week, on average. Toward the end of the study period, this content was shared by approximately 30 tweets per article per week, on average. However, as discussed in the main text, success is extremely heterogeneous across articles. This is the case irrespective of whether we measure success through the number of tweets (Fig.~\ref{fig:pop-distr}(a)) or accounts (Fig.~\ref{fig:pop-distr}(b)) sharing an article. For both popularity measures, the distributions are very broad and basically indistinguishable across articles from low-credibility vs. fact-checking sources.

\subsection*{Content Analysis}

Our analysis considers content published by a set of websites flagged as sources of misinformation by third-party journalistic and fact-checking organizations (Table~\ref{tab:sources}). This source-based approach relies on the assumption that most of the articles published by our compilation of sources are some type of misinformation, as we cannot fact-check each individual article. 
We validated this assumption by estimating the rate of false positives, i.e, verified articles, in the corpus. We manually evaluated a random sample of articles ($N=50$) drawn from our corpus, stratified by source. We considered only those sources whose articles were tweeted at least once in the period of interest. 
To draw an article, we first selected a source at random with replacement, and then chose one of the articles it published, again at random but without replacement. 
%To evaluate this sampling strategy, 
We repeated our analysis on an additional sample ($N=50$) 
%The first was a random sample in which each article had uniform chances of being drawn (`sample by article'), and thus corresponds to a `naive' sampling that is biased toward sources that produce a lot of articles. The second was a sample 
in which the chances of drawing an article are proportional to the number of times it was tweeted. This `sample by tweet' is thus biased toward more popular sources.

\begin{figure}
\centerline{\includegraphics[width=\bigfig]{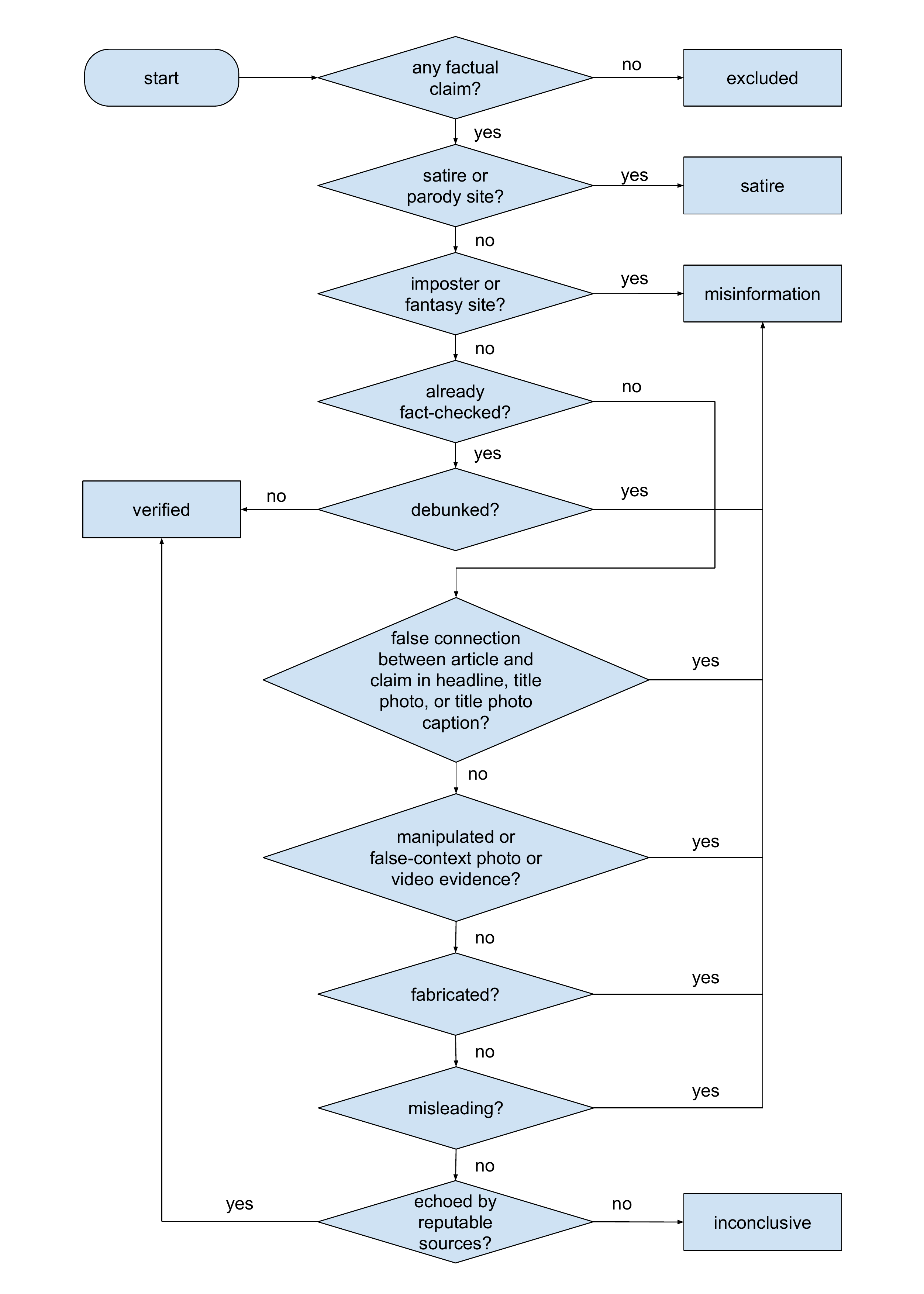}}
\caption{Flowchart summarizing the annotation rubric employed in the content analysis.}
\label{fig:flowchart}
\end{figure}

It is important to note that articles with unverified claims are sometimes updated after being debunked. This happens usually late, after the article has spread, and could lead to overestimating the rate of false positives. To mitigate this phenomenon, the earliest snapshot of each article was retrieved from the Wayback Machine at the Internet Archive (\url{archive.org}). If no snapshot was available, we retrieved the version of the page current at verification time. If the page was missing from the website or the website was down, we reviewed the title and body of the article crawled by Hoaxy. We gave priority to the current version over the possibly more accurate crawled version because, in deciding whether a piece of content is misinformation, we want to consider any form of visual evidence included with it, such as images or videos.

\begin{figure}
%\centerline{\includegraphics[width=\bigfig]{figs/sample_annotation_piechart.png}}
\centerline{\includegraphics[width=\bigfig]{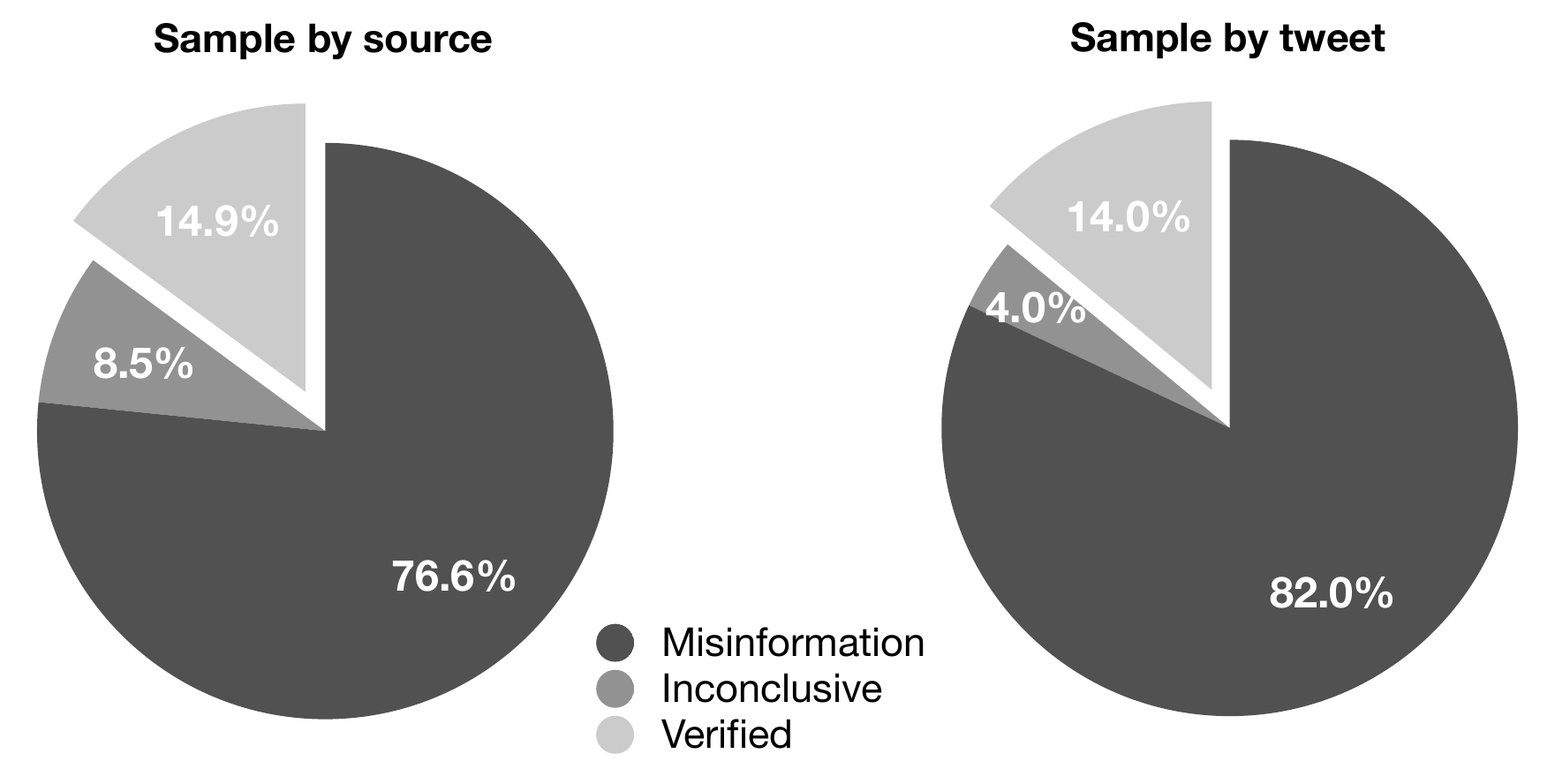}}
\caption{Content analysis based on two samples of articles. 
%The shaded area corresponds to the rate of false positives in our samples of claims. 
Sampling by source gives each source equal representation, while sampling by tweets 
%or by articles 
biases the analysis toward more popular 
%or prolific 
sources.
%, respectively.
We excluded from the sample by source three articles that did not contain any factual claims.
%We excluded any article with no factual claim. In the sample by article, source, and tweet there were six, three, and zero, such articles, respectively.
Satire articles are grouped with misinformation, as explained in the main text.
}
\label{fig:piechart}
\end{figure}

After retrieving all articles in the two samples, each article was evaluated independently by two reviewers (two of the authors), using a rubric summarized in Fig.~\ref{fig:flowchart}. Each article was then labeled with the majority label, with ties broken by a third reviewer (another author). Fig~\ref{fig:piechart} shows the results of the analysis. We report the fractions of articles that were verified and that could not be verified (inconclusive), out of the total number of articles that contain any factual claim. The rate of false positives is below 15\% in both samples.      

\subsection*{Concentration}

In the main text we use the Gini coefficient to calculate the concentration of posting activity for an article, based on the accounts that post links to the article. For each article, the Lorenz curve plots the cumulative share of tweets versus the cumulative share of accounts generating these tweets. The Gini coefficient is the ratio of the area that lies between the line of equality (diagonal) and the Lorenz curve, over the total area under the line of equality. A high coefficient indicates that a small subset of accounts was responsible for a large portion of the posts.

\subsection*{Bot Score Calibration}

Calibration methods are applicable when a machine learning classifier outputs probabilistic scores. Well-calibrated classifiers are probabilistic models for which the estimates can be directly interpreted as confidence levels. We use Platt's scaling~\cite{niculescu2005predicting}, a logistic regression model trained on classifier outputs, to calibrate the bot score computed by the Botometer classifier.

\begin{figure}
    \centering
    \includegraphics[width=\halffig]{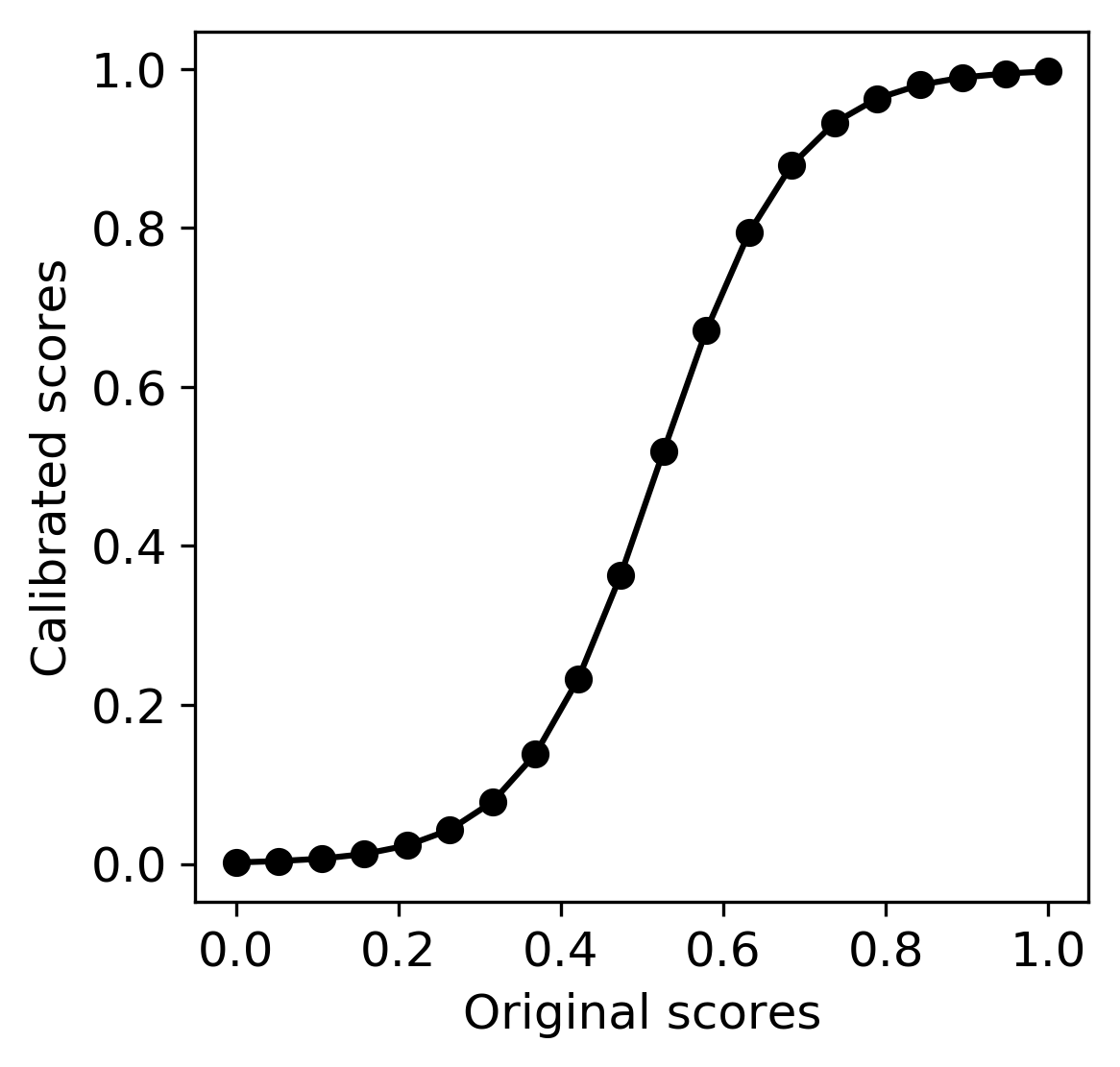}
    \includegraphics[width=\halffig]{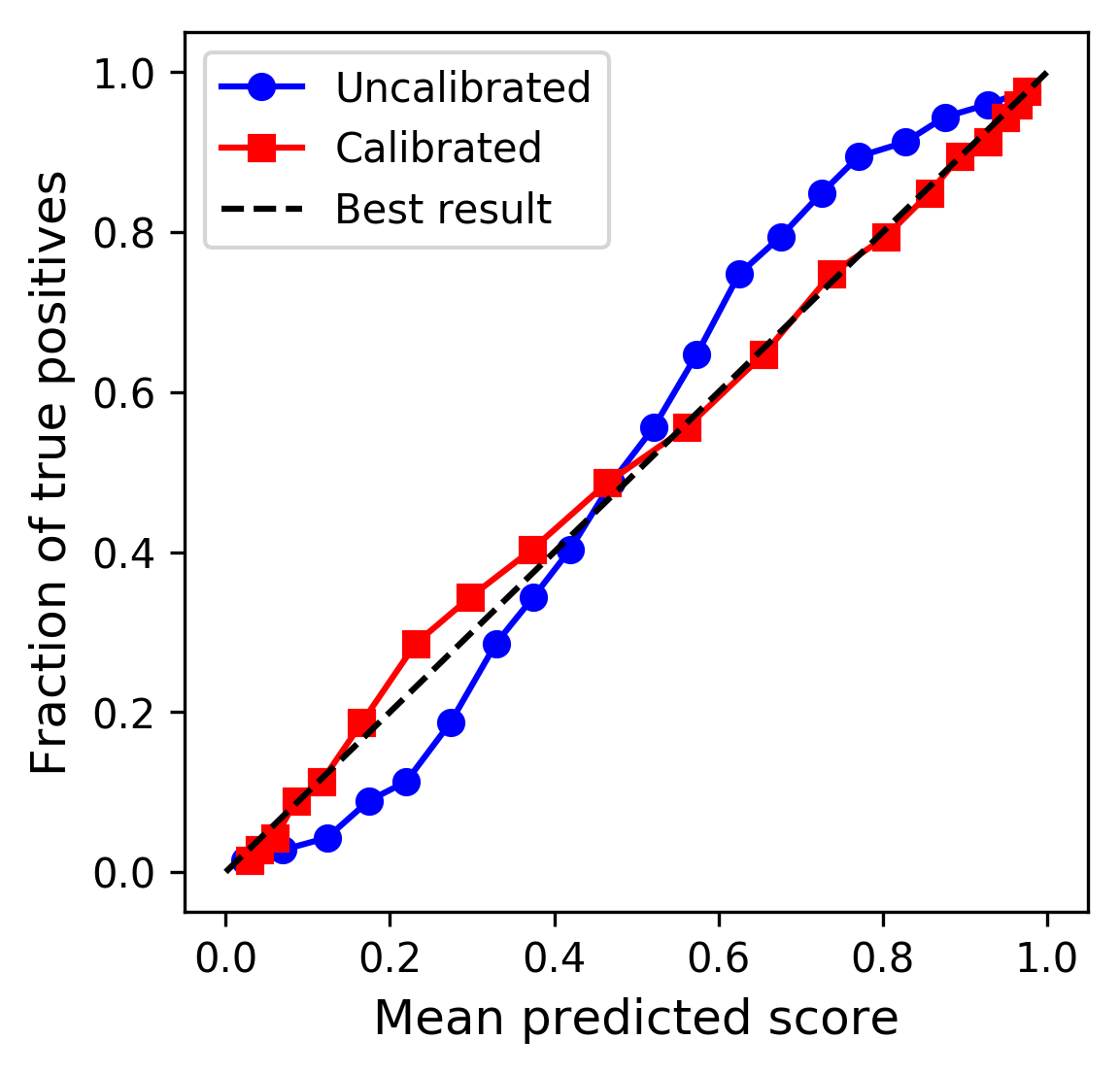}
    \caption{Bot score calibration curves. Left: The calibration mapping function projects raw classifier output to calibrated scores. Right: Reliability curves plot the true positive rate against the mean predicted scores. The calibrated curve indicates higher reliability because it is closer to the unbiased diagonal line.}
    \label{fig:botscore_calibration}
\end{figure}

We present the mapping between raw and calibrated scores in Fig.~\ref{fig:botscore_calibration}. The calibration only changes scores within the unit interval, but retains the ranking among accounts. The figure also shows reliability diagrams for raw and calibrated scores~\cite{degroot1983comparison}. We split the unit interval into 20 bins. Each instance in the training data set is assigned to a bin based on its predicted (raw) score. For each bin, the mean predicted score is computed and compared against the fraction of true positive cases. In a well-calibrated model, the points align to the diagonal. 

\subsection*{Bot Classification}

To show that a few social bots are disproportionately responsible for the spread of low-credibility content, we considered a random sample of accounts that shared at least one article from a low-credibility source, and evaluated these accounts using the bot classification system Botometer. Out of 1,000 sampled accounts, 85 could not be inspected because they had been either suspended, deleted, or turned private. For each of the remaining 915, Botometer returned a \emph{bot score} estimating the level of automation of the account. To quantify how many accounts are likely bots, we transform bot scores into binary assessments using a threshold of 0.5. This is a conservative choice to minimize false negatives and especially false positives, as shown in prior work (cit. in main text). Table~\ref{tab:botscores} shows the fraction of accounts with scores above the threshold. To give a sense of their overall impact in the spreading of low-credibility content, Table~\ref{tab:botscores} also shows the fraction of tweets with articles from low-credibility sources posted by accounts that are likely bots, and the number of unique articles included in those tweets overall. As a comparison, we also tally the fact-checks shared by these accounts, showing that bot accounts focused on sharing low-credibility content and ignored fact-checking content.

\begin{table}
    \caption{Analysis of likely bots and their content spreading activity based on a random sample of Twitter accounts sharing at least one article from a low-credibility source.}
    \centering
    \begin{tabular}{lrrr}
    \hline
    & Total & Likely bots & Percentage \\
    \hline
    Accounts & 915 & 54 & 6\% \\
    Tweets with low-credibility articles & 11,656 & 3,587 & 31\% \\
    Unique low-credibility articles & 7,726 & 2,608 & 34\% \\
    Tweets with fact-checks & 598 & 4 & 0.7\% \\
    Unique fact-checks & 395 & 3 & 0.8\% \\
    \hline
    \end{tabular}
    \label{tab:botscores}
\end{table}

In the main text we show the distributions of bot scores for this sample of accounts, as well as for a sample of accounts that have been most active in spreading low-credibility content (\emph{super-spreaders}). To select the super-spreaders, we ranked all accounts by how many tweets they posted with links to low-credibility sources, and considered the top 1,000 accounts. We then performed the same classification steps discussed above. For the same reasons mentioned above, we could not obtain scores for 39 of these accounts, leaving us with a sample of 961 scored accounts.

\section*{Appendix: Supplementary Text} 

\subsection*{Super-Spreaders of Low-Credibility Content}

\begin{figure}
\centering
\includegraphics[width=\smallfig,valign=t]{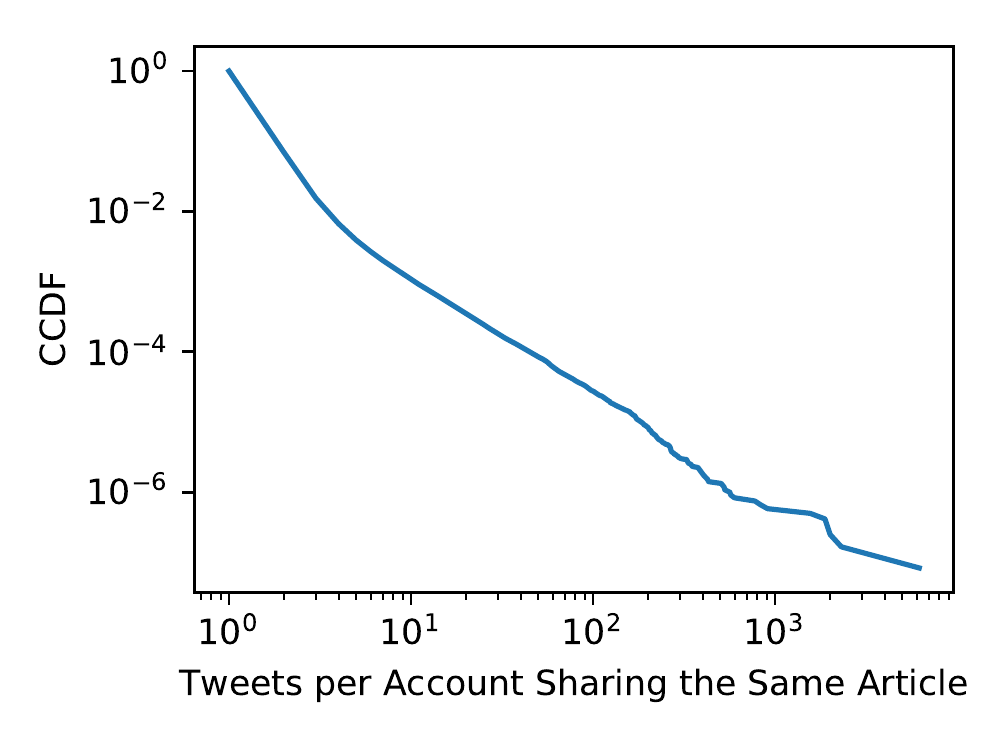}
\caption{Cumulative distribution of repetitions, i.e., the number of times a single account tweets the same link to an article from a low-credibility source.}
\label{fig:repetitions}
\end{figure}

In the main text we show that the more popular a low-credibility article, the more its posting activity is concentrated around a relative small number of active accounts. We also find that the most active spreaders of content from low-credibility sources are more likely to be social bots. To further illustrate the anomalous activity patterns of these ``super-spreaders,'' Fig.~\ref{fig:repetitions} plots the distribution of repeated tweets by individual accounts sharing the same low-credibility article. While it is normal behavior for a person to share an article once, the long tail of the distribution highlights inorganic, automated support. A single account posting the same article over and over --- hundreds or thousands of times in some cases --- is likely controlled by software. 

\subsection*{Bots Targeting Influentials}

\begin{figure}
\centerline{\includegraphics[width=\bigfig]{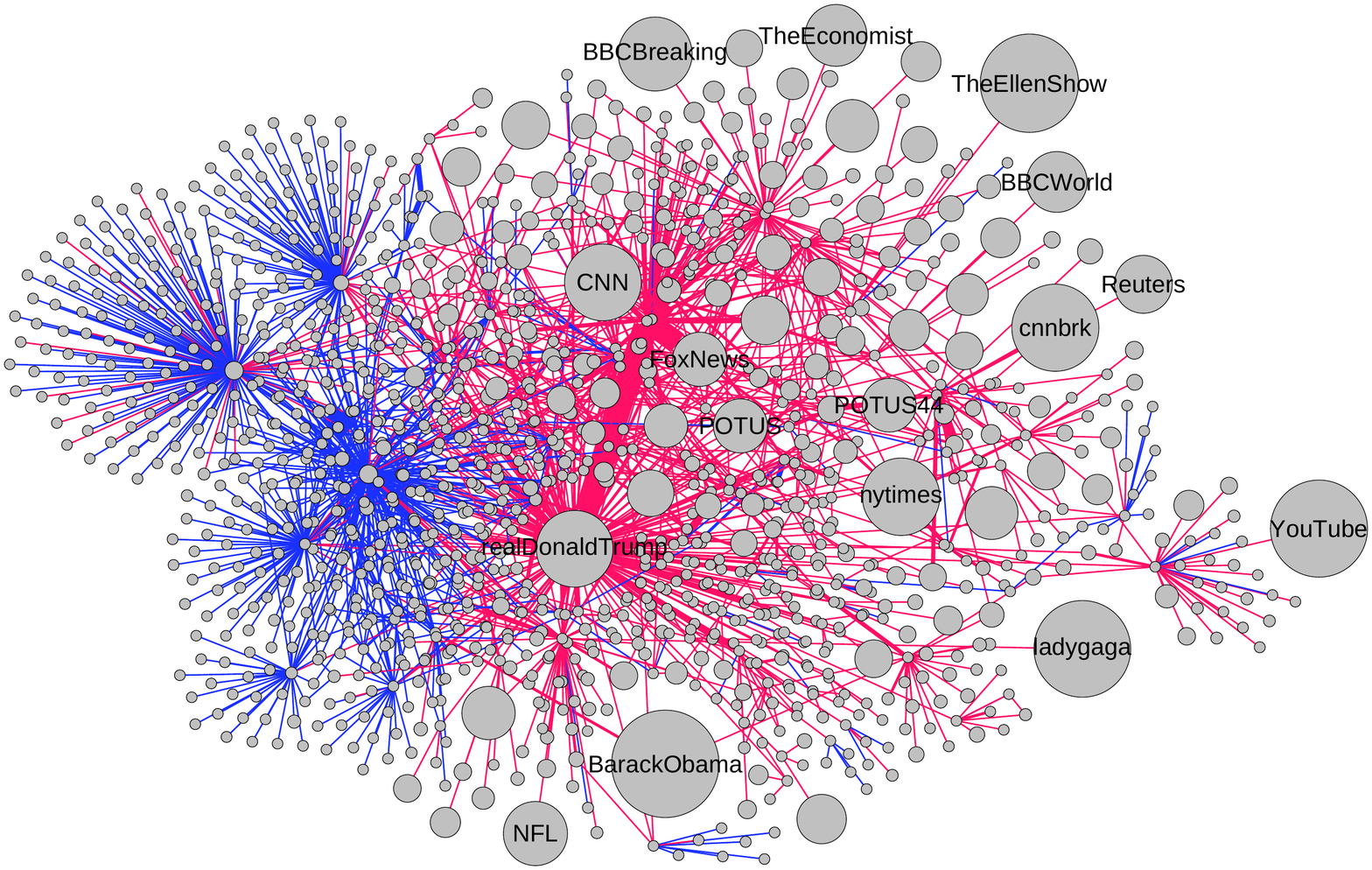}}
\caption{Example of targeting for the article \textit{Report: three million votes in presidential election cast by illegal aliens}, published by \protect\url{Infowars.com} on November 14, 2016 and shared over 18 thousand times on Twitter. Only a portion of the diffusion network is shown. Nodes stand for Twitter accounts, with size representing number of followers. Links illustrate how the article spreads: by retweets and quoted tweets (blue), or by replies and mentions (red).}
\label{fig:bots-targeting-influentials}
\end{figure}

The main text discusses a strategy used by bots, by which influential users are mentioned in tweets that link to low-credibility content. Bots seem to employ this targeting strategy repetitively. Fig.~\ref{fig:bots-targeting-influentials} offers an illustration: in this example, a (now suspended) single account produced 19 tweets linking to the article shown in the figure and mentioning \texttt{@realDonaldTrump}.  

\subsection*{Amplification by Bots}

\begin{figure}
    \centering
    \includegraphics[width=\bigfig]{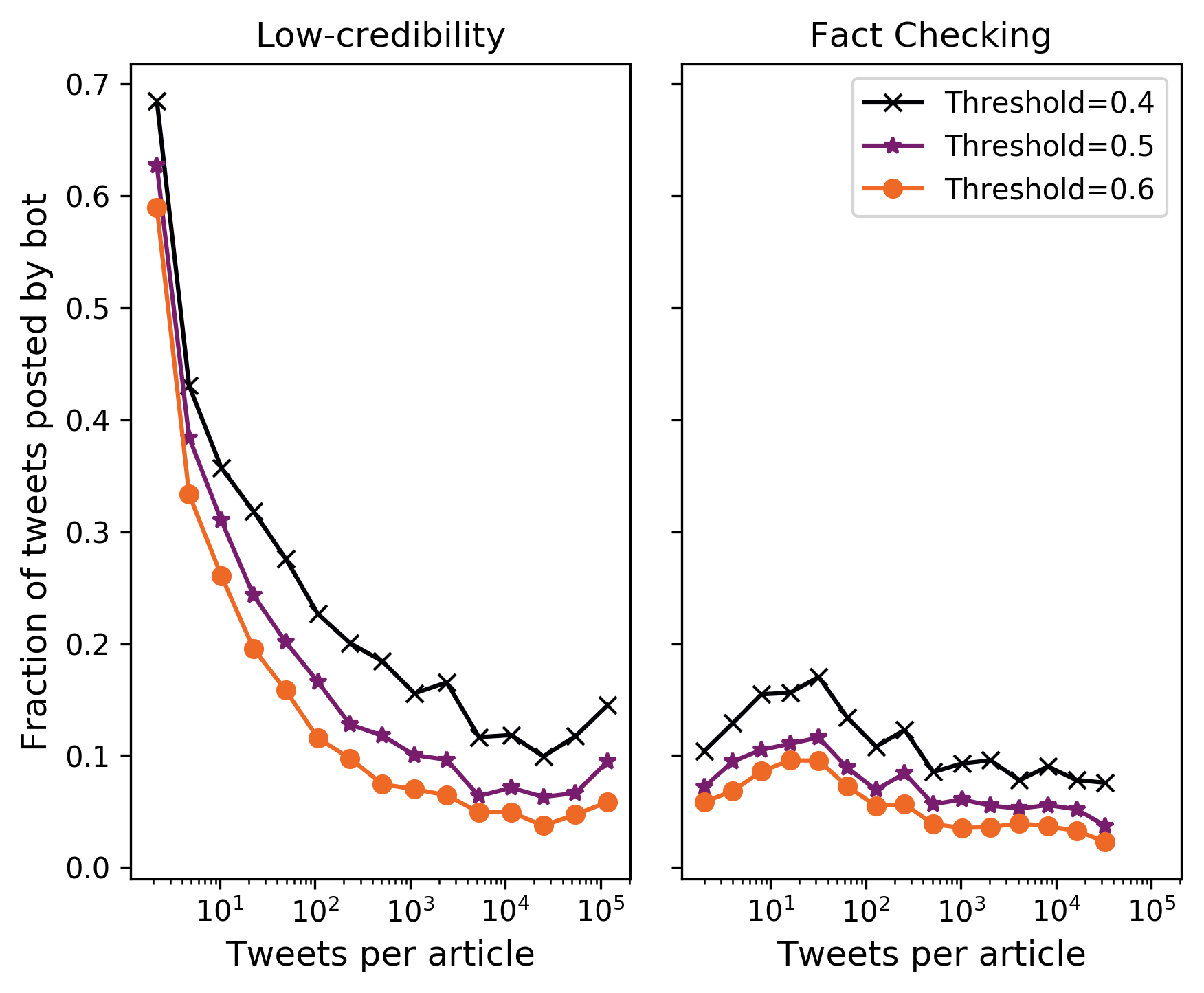}
    \caption{Fraction of tweets linking to news articles that are posted by accounts with bot score above a threshold, as a function of the popularity of the linked articles. We see different bot activity for articles from low-credibility (left) versus fact-checking (right) sources.}
    \label{fig:bot_fraction_estimate}
\end{figure}

The analysis in the main text focuses on the role of bots in the spread of articles from low-credibility sources, assuming that bots do not equally support the spread of articles from fact-checking sources. In fact, we show in the main text that articles from low-credibility and fact-checking sources spread through different mixes of original tweets, retweets, and replies. And we also find that low-credibility sources have greater support from bots than fact-checking and satire sources. To further confirm the assumption that bots do not play an equal role in the spread of fact-checking articles, we observe in Fig.~\ref{fig:bot_fraction_estimate} that the fraction of tweets posted by likely bots is much higher for articles from low-credibility sources. Further, the fraction depends on popularity in the case of articles from low-credibility sources (it gets diluted for more viral ones), whereas it is flatter for articles from fact-checking sources. Here, bots and humans are separated based on a threshold in the bot score. These findings are robust to the choice of threshold, and point to selective amplification of articles from low-credibility sources by bots.

\begin{figure}
    \centering
    \includegraphics[width=\bigfig]{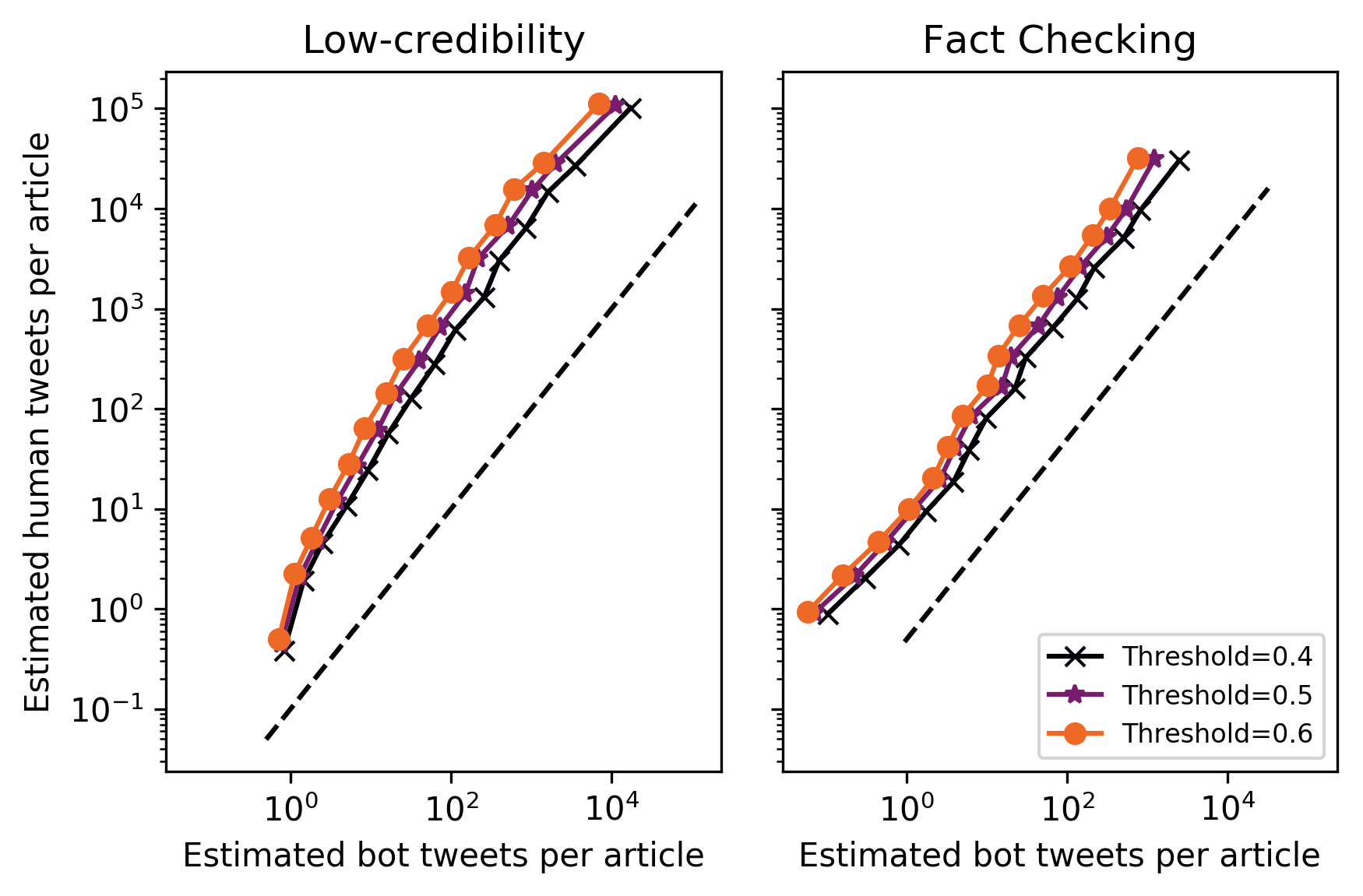}
    \caption{For links to articles from low-credibility (left) and fact-checking (right) sources, the number of tweets by accounts with bot score above a threshold is plotted versus the number of tweets by accounts with bot score below the threshold. The dashed lines are guides to the eye, showing linear growth. A super-linear relationship is a signature of amplification by bots.}
    \label{fig:bot_amplification}
\end{figure}

To focus on amplification more directly, let us consider how exposure to humans varies with activity by bots. Fig.~\ref{fig:bot_amplification} estimates the numbers of tweets by likely humans/bots, using a threshold on bot scores to separate them. Results are robust with respect to the choice of threshold. For articles from low-credibility sources, the estimated number of human tweets per article grows faster than the estimated number of bot tweets for article. For fact-checking articles, instead, we find a linear relationship. In other words, bots seem to amplify the reach of articles from low-credibility sources, but not the reach of articles from fact-checking sources. 

\subsection*{Geographic Targeting}

\begin{figure}
\centerline{\includegraphics[width=\bigfig]{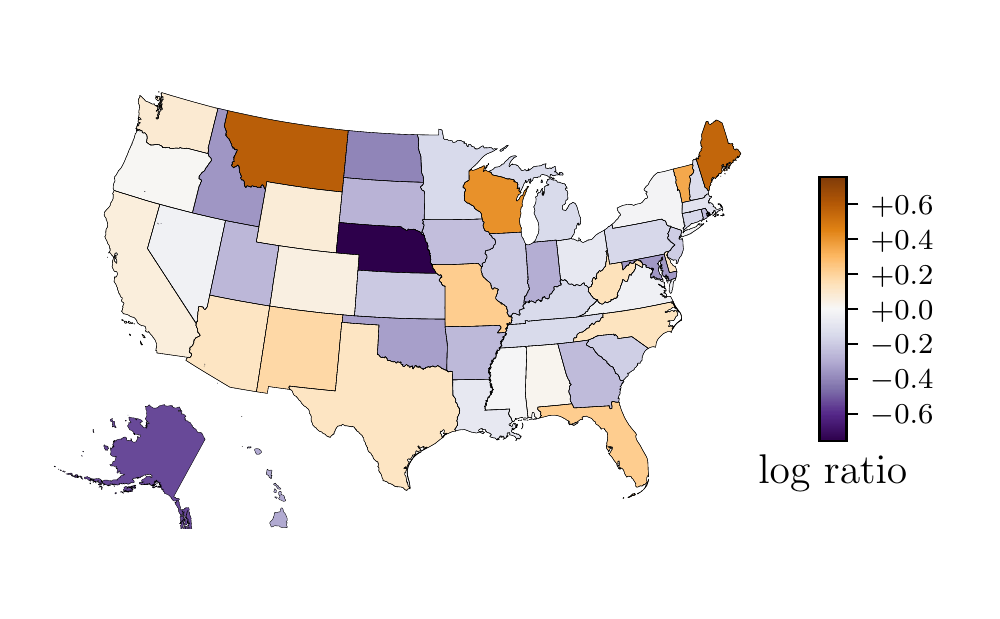}}
\caption{Map of location targeting by bots, relative to a baseline. To gauge sharing activity by likely bots, we considered 793,909 tweets in the period between August and October 2016 posting links to low-credibility articles by accounts with bot score above 0.5 that reported a U.S. state location in their profile. We compared the tweet frequencies by states with those expected from a sample of 1,393,592,062 tweets in the same period, obtained from a 10\% random sample of public posts from the Twitter streaming API. Colors indicate log ratios after normalization: positive values indicate states with higher than expected bot activity. Using different thresholds yields similar maps.}
\label{fig:bots-map}
\end{figure}

We examined whether bots (or rather their programmers) tended to target voters in certain states by creating the appearance of users posting from those locations. To this end, we considered accounts with bot scores above a threshold that shared articles from low-credibility sources in the three months before the election, and focused on those accounts with a state location in their profile. The location is self-reported and thus trivial to fake. We compared the distribution of bot account locations across states with a baseline obtained from a large sample of tweets in the same period. A $\chi^2$ test indicates that the location patterns produced by bots are inconsistent with the geographic distribution of conversations on Twitter ($p<10^{-4}$). This suggests that as part of their disguise, social bots are more likely to report certain locations than others. For example, Fig.~\ref{fig:bots-map} shows higher than expected bot activity appearing to originate from the states of Montana, Maine, and Wisconsin. However, we did not find evidence that bots used this particular strategy for geographic targeting of swing states (Table~\ref{tab:spearman}).

\begin{table}
    \caption{Spearman rank correlation $\rho$ between U.S. state absolute expected vote margin and relative bot activity. Low margins indicate 2016 tight races (swing states). Data from FiveThirtyEight (\url{projects.fivethirtyeight.com/2016-election-forecast/}). Bot activity is measured by numbers of tweets posting links to low-credibility articles by accounts with bot score above threshold that reported a U.S. state location in their profile, in the period between August and October 2016. The counts by states are normalized by those expected from a sample of 1,393,592,062 tweets in the same period, obtained from a 10\% random sample of public posts from the Twitter streaming API. The positive correlation indicates that bot activity was not associated with swing states. The $p$-values are based on two-tailed t-tests.}
    \centering
    \begin{tabular}{crcc}
    \hline
    Bot score threshold & Tweets by likely bots & $\rho$ & $p$-value \\
    \hline
    0.4 & 1,135,816 & 0.28 & 0.04 \\
    0.5 &   793,909 & 0.29 & 0.04 \\
    0.6 &   588,751 & 0.30 & 0.03 \\
    \hline
    \end{tabular}
    \label{tab:spearman}
\end{table}

\subsection*{Robustness Analyses}

The results in the main text are robust with respect to various choices and assumptions, presented next. 

\subsubsection*{Criteria for selection of sources}

\begin{figure}
\centerline{\includegraphics[width=\bigfig]{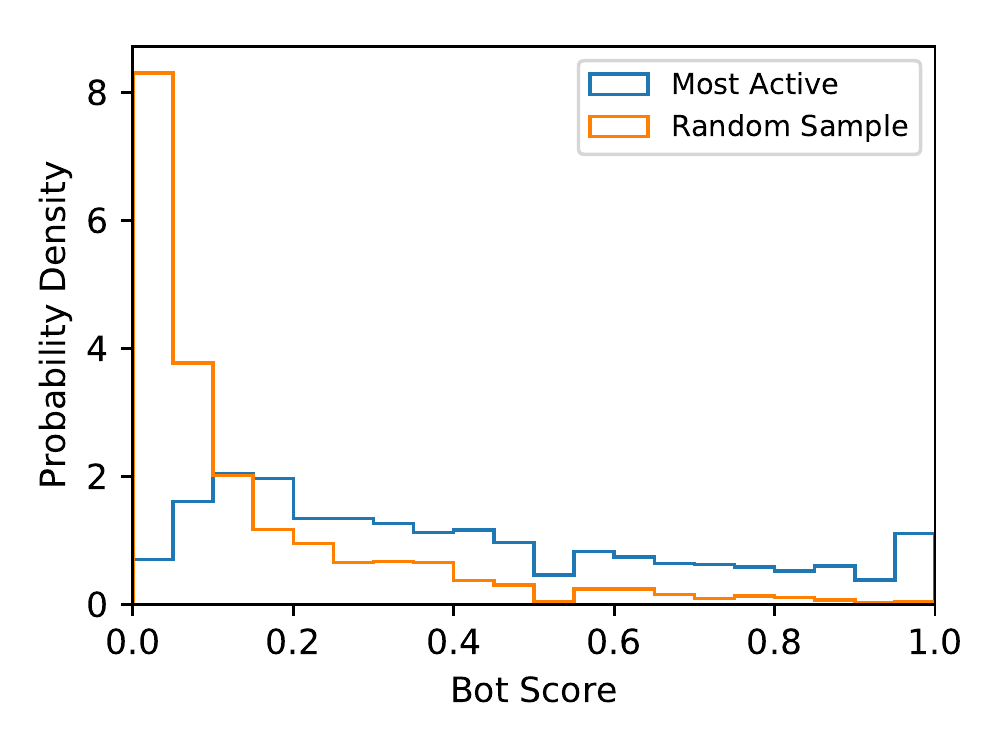}}
\caption{Bot score distributions for super-spreaders vs. randomly selected sharers of links to low-credibility sources selected by the consensus criterion. 
% * Fig4. 1000 accounts were randomly sampled and 1000 most active (in terms of tweets posted) accounts were collected as a comparison.
The random sample includes 992 accounts who posted at least one link to an article from a low-credibility source. Their bot scores are compared to 997 accounts that most actively share such links. 
% Mann Whitney U test: statistic=754575, p=1.03e-130. 
The two groups have significantly different scores ($p<10^{-4}$ according to a Mann-Whitney $U$ test). 
% THE NUMBERS BELOW ARE PRE-CALIBRATION
% 7\% of accounts in the random sample and 37\% of accounts in the most active group have bot score above 0.5.
}
\label{fig:supp-bots-of-users}
\end{figure}

\begin{figure}
\centerline{\includegraphics[width=\bigfig]{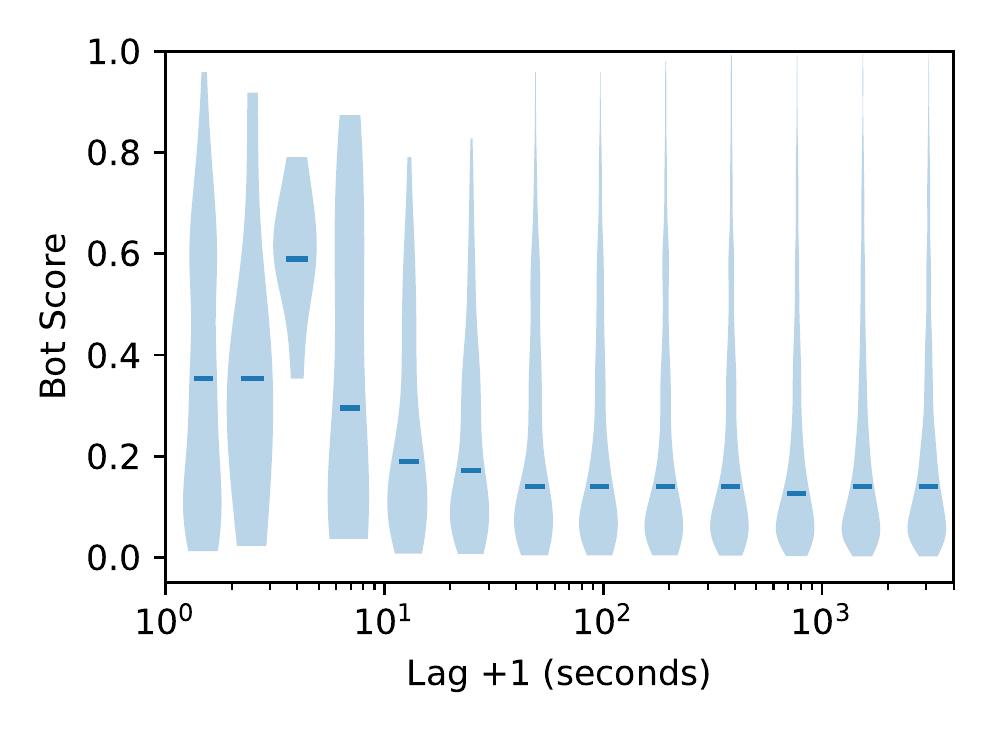}}
\caption{Temporal evolution of bot support after the first share of a viral story from a consensus low-credibility source. We consider a random sample of 20,000 accounts out of the 163,563 accounts that participate in the spread of the 1,000 most viral articles. 
After articles from \textit{The Onion} are excluded, we are left with 42,202 tweets from 13,926 accounts. 
We align the times when each link first appears. We focus on a one-hour early spreading phase following each of these events, and divide it into logarithmic lag intervals.
The plot shows the bot score distribution for accounts sharing the links during each of these lag intervals. 
}
\label{fig:supp-bots-in-early-spreading}
\end{figure}

\begin{figure}
\centerline{\includegraphics[width=\mediumfig]{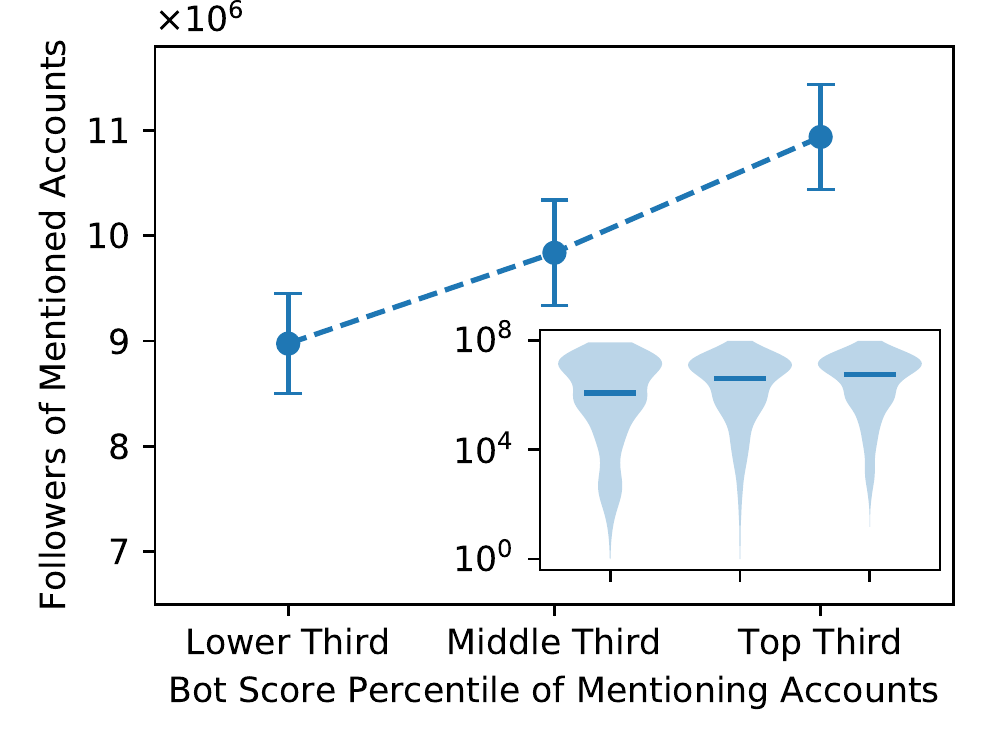}}
\caption{Average number of followers for Twitter users who are mentioned (or replied to) by a sample of 20,000 accounts that link to the 1,000 most viral articles from consensus low-credibility sources. 
% started from 33,333 pairs of (mentioning_id, mentioned_id), 4061 unique mentioning_id and 5022 unique mentioned_id
We obtained bot scores for 4,006 unique mentioning accounts and 4,965 unique mentioned accounts, participating in 33,112 mention/reply pairs. We excluded 13,817 of these pairs using the ``via @screen\_name'' mentioning pattern.
The mentioning accounts are aggregated into three groups by bot score percentile. Error bars indicate standard errors. Inset: Distributions of follower counts for users mentioned by accounts in each percentile group.}
\label{fig:supp-bots-targeting-influentials}
\end{figure}

\begin{figure}
\centerline{\includegraphics[width=\textwidth]{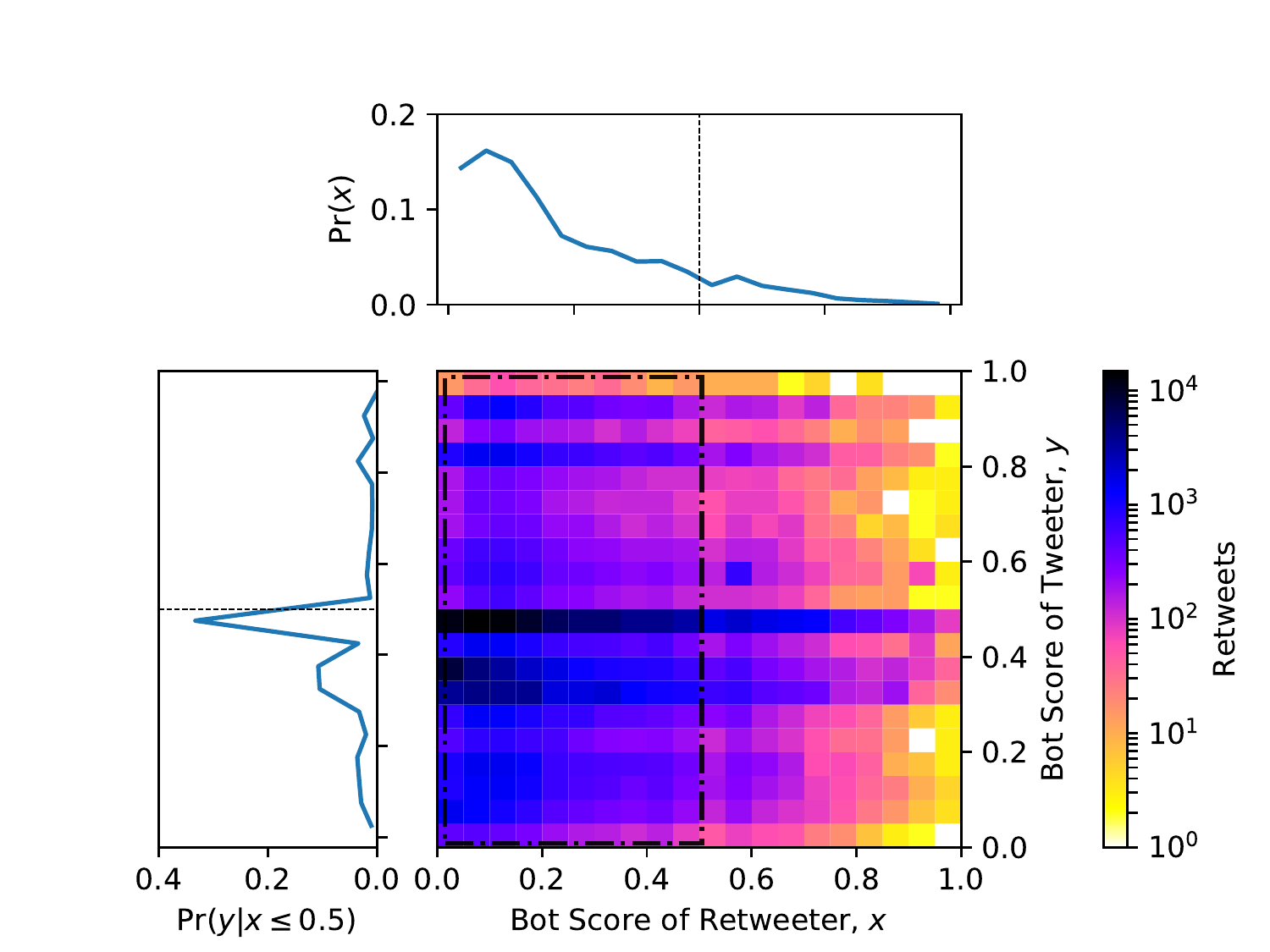}}
\caption{Joint distribution of the bot scores of accounts that retweeted links to articles from consensus low-credibility sources and accounts that had originally posted the links. We considered retweets by a sample of 20,000 accounts that posted the 1,000 most viral articles. 
% We got 231327 retweets, in which  18,255 unique retweeter_id and 12,931 unique tweeter_id.
We obtained bot scores for 12,792 tweeting accounts and 17,664 retweeting accounts, participating in 229,725 retweet pairs. 
Color represents the number of retweeted messages in each bin, on a log scale. Projections show the distributions of bot scores for retweeters (top) and for accounts retweeted by likely humans (left).}
\label{fig:supp-bots-retweeter-vs-tweeter}
\end{figure}

We repeated the analyses in the main text using the more restrictive criterion for selecting low-credibility sources, based on a consensus among three or more news and fact-checking organizations. The 65 consensus sources are listed in Table~\ref{tab:sources}. To carry out these analyses, we inspected 33,115 accounts and could obtain bot scores for 32,250 of them; the rest had been suspended or gone private. The results are qualitatively similar to those in the main text and support the robustness of the findings, namely: super-spreaders of articles from low-credibility sources are likely bots (Fig.~\ref{fig:supp-bots-of-users}), bots amplify the spread of information from low-credibility sources in the early phases (Fig.~\ref{fig:supp-bots-in-early-spreading}), bots target influential users (Fig.~\ref{fig:supp-bots-targeting-influentials}), and humans retweet low-credibility content posted by bots (Fig.~\ref{fig:supp-bots-retweeter-vs-tweeter}).

\subsubsection*{Absence of correlation between activity and bot score}

\begin{figure}
    \centering
    \includegraphics{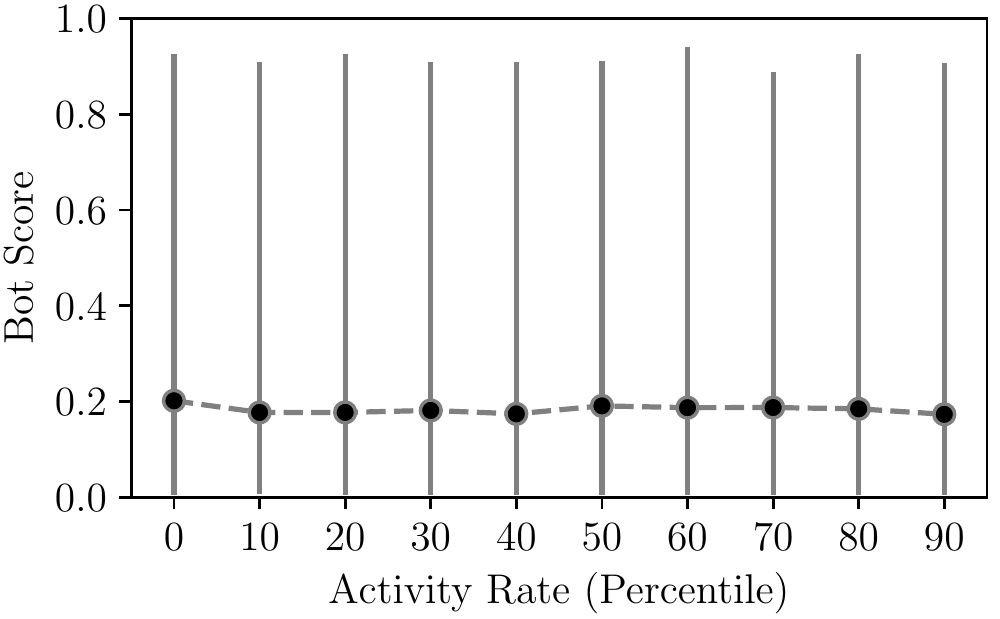}
    \caption{Distributions of bot scores versus account activity. 
    For this analysis we randomly selected 48,517 distinct Twitter accounts evaluated by Botometer. Of these, 11,190 were available for crawling their profiles and measuring their activity (number of tweets). Bins correspond to deciles in the activity rate. 
    We show the average and 95\% confidence interval for the bot score distribution of the accounts in each activity bin. 
    There is no correlation between activity and bot score (Pearson's $\rho = -0.007$).}
    \label{fig:activity_vs_botscore}
\end{figure}

Our notion of super-spreader is based upon ranking accounts by activity and taking those above a threshold. The analysis about super-spreaders of low-credibility content being likely bots assumes that this finding is not explained by a correlation between activity and bot score. In fact, although the bot classification model does consider volume of tweets as one among over a thousand features, it is not trained in such a way that there is an obvious monotonic relation between activity and bot score. A simple monotonic relation between overall volume and bot score would lead to many false positives, because many bots produce very few tweets or appear to produce none (they delete their tweets); these accounts still get high bot scores. Figure~\ref{fig:activity_vs_botscore} confirms that account activity volume and bot scores are uncorrelated.

\subsubsection*{Bot-score threshold values}

The results are also not affected by the use of different bot-score thresholds to separate social bots and human accounts. For example, we experimented with different thresholds and found that they do not change our conclusions that super-spreaders are more likely to be social bots. Figs.~\ref{fig:bot_fraction_estimate} and \ref{fig:bot_amplification} and Table~\ref{tab:spearman} show that other findings are also robust with respect to the bot score threshold, even though the estimated percentages of likely humans/bots, and the estimated numbers of tweets posted by them, are naturally sensitive to the threshold.

% References
\bibliographystyle{abbrv}
\bibliography{refs}

\begin{thebibliography}{10}

\bibitem{albert2000error}
R.~Albert, H.~Jeong, and A.-L. Barab{\'a}si.
\newblock Error and attack tolerance of complex networks.
\newblock {\em Nature}, 406(6794):378--382, 2000.

\bibitem{allcott2017social}
H.~Allcott and M.~Gentzkow.
\newblock Social media and fake news in the 2016 election.
\newblock {\em Journal of Economic Perspectives}, 31(2):211--236, 2017.

\bibitem{10.1371/journal.pone.0118093}
A.~Bessi, M.~Coletto, G.~A. Davidescu, A.~Scala, G.~Caldarelli, and
  W.~Quattrociocchi.
\newblock Science vs conspiracy: Collective narratives in the age of
  misinformation.
\newblock {\em PLoS ONE}, 10(2):1--17, 2015.

\bibitem{bessi2016social}
A.~Bessi and E.~Ferrara.
\newblock Social bots distort the 2016 us presidential election online
  discussion.
\newblock {\em First Monday}, 21(11), 2016.

\bibitem{boshmaf2011socialbot}
Y.~Boshmaf, I.~Muslukhov, K.~Beznosov, and M.~Ripeanu.
\newblock The socialbot network: when bots socialize for fame and money.
\newblock In {\em Proceedings of the 27th annual ACM computer security
  applications conference}, pages 93--102, 2011.

\bibitem{Castillo2011}
C.~Castillo, M.~Mendoza, and B.~Poblete.
\newblock Information credibility on {T}witter.
\newblock In {\em Proceedings of the 20th International Conference on World
  Wide Web}, page 675, 2011.

\bibitem{chu2010tweeting}
Z.~Chu, S.~Gianvecchio, H.~Wang, and S.~Jajodia.
\newblock Who is tweeting on twitter: human, bot, or cyborg?
\newblock In {\em Proceedings of the 26th annual ACM computer security
  applications conference}, pages 21--30, 2010.

\bibitem{Truthy_icwsm2011politics}
M.~Conover, J.~Ratkiewicz, M.~Francisco, B.~Gon\c{c}alves, A.~Flammini, and
  F.~Menczer.
\newblock Political polarization on twitter.
\newblock In {\em Proc. 5th International AAAI Conference on Weblogs and Social
  Media (ICWSM)}, 2011.

\bibitem{conover12partisan}
M.~D. Conover, B.~Gon\c{c}alves, A.~Flammini, and F.~Menczer.
\newblock Partisan asymmetries in online political activity.
\newblock {\em EPJ Data Science}, 1:6, 2012.

\bibitem{Davis16BotOrNot}
C.~A. Davis, O.~Varol, E.~Ferrara, A.~Flammini, and F.~Menczer.
\newblock Botornot: A system to evaluate social bots.
\newblock In {\em Proceedings of the 25th International Conference Companion on
  World Wide Web}, pages 273--274, 2016.

\bibitem{degroot1983comparison}
M.~H. DeGroot and S.~E. Fienberg.
\newblock The comparison and evaluation of forecasters.
\newblock {\em The statistician}, pages 12--22, 1983.

\bibitem{del2016spreading}
M.~Del~Vicario, A.~Bessi, F.~Zollo, F.~Petroni, A.~Scala, G.~Caldarelli, H.~E.
  Stanley, and W.~Quattrociocchi.
\newblock The spreading of misinformation online.
\newblock {\em Proc. National Academy of Sciences}, 113(3):554--559, 2016.

\bibitem{ferrara2017frenchbots}
E.~Ferrara.
\newblock {Disinformation and Social Bot Operations in the Run Up to the 2017
  French Presidential Election}.
\newblock {\em First Monday}, 22(8), 2017.

\bibitem{socialbots-CACM}
E.~Ferrara, O.~Varol, C.~Davis, F.~Menczer, and A.~Flammini.
\newblock The rise of social bots.
\newblock {\em Comm. ACM}, 59(7):96--104, 2016.

\bibitem{Pew2016}
J.~Gottfried and E.~Shearer.
\newblock News use across social media platforms 2016.
\newblock White paper, Pew Research Center, May 2016.

\bibitem{TrendMicro}
L.~Gu, V.~Kropotov, and F.~Yarochkin.
\newblock The fake news machine: How propagandists abuse the internet and
  manipulate the public.
\newblock Trendlabs research paper, Trend Micro, 2017.

\bibitem{guess2018selective}
A.~Guess, B.~Nyhan, and J.~Reifler.
\newblock {Selective Exposure to Misinformation: Evidence from the consumption
  of fake news during the 2016 US presidential campaign}.
\newblock Unpublished manuscript, 2018.

\bibitem{gupta2014tweetcred}
A.~Gupta, P.~Kumaraguru, C.~Castillo, and P.~Meier.
\newblock Tweetcred: Real-time credibility assessment of content on twitter.
\newblock In {\em International Conference on Social Informatics}, pages
  228--243, 2014.

\bibitem{Hassan2014}
N.~Hassan, A.~Sultana, Y.~Wu, G.~Zhang, C.~Li, J.~Yang, and C.~Yu.
\newblock Data in, fact out: Automated monitoring of facts by factwatcher.
\newblock {\em Proc. VLDB Endow.}, 7(13):1557--1560, Aug. 2014.

\bibitem{Hodas12socialcom}
N.~O. Hodas and K.~Lerman.
\newblock How limited visibility and divided attention constrain social
  contagion.
\newblock In {\em Proc. ASE/IEEE International Conference on Social Computing},
  2012.

\bibitem{10.1371/journal.pmed.1002153}
P.~J. Hotez.
\newblock Texas and its measles epidemics.
\newblock {\em PLOS Medicine}, 13(10):1--5, 10 2016.

\bibitem{Forum:2013}
L.~Howell et~al.
\newblock Digital wildfires in a hyperconnected world.
\newblock In {\em Global Risks}. World Economic Forum, 2013.

\bibitem{Menczer06socialphishing}
T.~Jagatic, N.~Johnson, M.~Jakobsson, and F.~Menczer.
\newblock Social phishing.
\newblock {\em Communications of the ACM}, 50(10):94--100, October 2007.

\bibitem{jin2014news}
Z.~Jin, J.~Cao, Y.-G. Jiang, and Y.~Zhang.
\newblock News credibility evaluation on microblog with a hierarchical
  propagation model.
\newblock In {\em Proc. IEEE International Conference on Data Mining (ICDM)},
  pages 230--239, 2014.

\bibitem{Jun06062017}
Y.~Jun, R.~Meng, and G.~V. Johar.
\newblock Perceived social presence reduces fact-checking.
\newblock {\em Proceedings of the National Academy of Sciences},
  114(23):5976--5981, 2017.

\bibitem{KahanIdeology2013}
D.~M. Kahan.
\newblock Ideology, motivated reasoning, and cognitive reflection.
\newblock {\em Judgment and Decision Making}, 8(4):407--424, 2013.

\bibitem{fake-news-manifesto}
D.~Lazer, M.~Baum, Y.~Benkler, A.~Berinsky, K.~Greenhill, F.~Menczer,
  M.~Metzger, B.~Nyhan, G.~Pennycook, D.~Rothschild, M.~Schudson, S.~Sloman,
  C.~Sunstein, E.~Thorson, D.~Watts, and J.~Zittrain.
\newblock The science of fake news.
\newblock {\em Science}, 359(6380):1094--1096, 2018.

\bibitem{lee2010uncovering}
K.~Lee, J.~Caverlee, and S.~Webb.
\newblock Uncovering social spammers: social honeypots+ machine learning.
\newblock In {\em Proceedings of the 33rd international ACM SIGIR conference on
  Research and development in information retrieval}, pages 435--442, 2010.

\bibitem{Levendusky2013}
M.~S. Levendusky.
\newblock Why do partisan media polarize viewers?
\newblock {\em American Journal of Political Science}, 57(3):611–--623, 2013.

\bibitem{LEWANDOWSKY2017353}
S.~Lewandowsky, U.~K. Ecker, and J.~Cook.
\newblock Beyond misinformation: Understanding and coping with the
  ``post-truth'' era.
\newblock {\em Journal of Applied Research in Memory and Cognition},
  6(4):353--369, 2017.

\bibitem{Lippmann1922}
W.~Lippmann.
\newblock {\em Public Opinion}.
\newblock Harcourt, Brace and Company, 1922.

\bibitem{liu2015real}
X.~Liu, A.~Nourbakhsh, Q.~Li, R.~Fang, and S.~Shah.
\newblock Real-time rumor debunking on twitter.
\newblock In {\em Proceedings of the 24\textsuperscript{th} ACM International
  on Conference on Information and Knowledge Management}, CIKM '15, pages
  1867--1870, New York, NY, USA, 2015. ACM.

\bibitem{Markines09airweb}
B.~Markines, C.~Cattuto, and F.~Menczer.
\newblock Social spam detection.
\newblock In {\em Proc. 5th International Workshop on Adversarial Information
  Retrieval on the Web (AIRWeb)}, 2009.

\bibitem{metaxas2015using}
P.~T. Metaxas, S.~Finn, and E.~Mustafaraj.
\newblock Using twittertrails.com to investigate rumor propagation.
\newblock In {\em Proceedings of the 18th ACM Conference Companion on Computer
  Supported Cooperative Work \& Social Computing}, CSCW'15 Companion, pages
  69--72, 2015.

\bibitem{FB-LOW-Q}
A.~Mosseri.
\newblock News feed fyi: Showing more informative links in news feed.
\newblock press release, Facebook, June 2017.

\bibitem{mustafaraj2010obscurity}
E.~Mustafaraj and P.~T. Metaxas.
\newblock From obscurity to prominence in minutes: Political speech and
  real-time search.
\newblock In {\em Proc. Web Science Conference: Extending the Frontiers of
  Society On-Line}, 2010.

\bibitem{Azadeh_pop_2017}
A.~Nematzadeh, G.~L. Ciampaglia, F.~Menczer, and A.~Flammini.
\newblock How algorithmic popularity bias hinders or promotes quality.
\newblock Preprint 1707.00574, arXiv, 2017.

\bibitem{niculescu2005predicting}
A.~Niculescu-Mizil and R.~Caruana.
\newblock Predicting good probabilities with supervised learning.
\newblock In {\em Proceedings of the 22nd International Conference on Machine
  Learning}, pages 625--632, 2005.

\bibitem{Nikolov15socialbubbles}
D.~Nikolov, D.~F.~M. Oliveira, A.~Flammini, and F.~Menczer.
\newblock Measuring online social bubbles.
\newblock {\em PeerJ Computer Science}, 1(e38), 2015.

\bibitem{Pariser}
E.~Pariser.
\newblock {\em The filter bubble: How the new personalized Web is changing what
  we read and how we think}.
\newblock Penguin, 2011.

\bibitem{low_quality_nhb2017}
X.~Qiu, D.~F.~M. Oliveira, A.~S. Shirazi, A.~Flammini, and F.~Menczer.
\newblock Limited individual attention and online virality of low-quality
  information.
\newblock {\em Nature Human Behavior}, 1:0132, 2017.

\bibitem{Truthy_icwsm2011class}
J.~Ratkiewicz, M.~Conover, M.~Meiss, B.~Gon\c{c}alves, A.~Flammini, and
  F.~Menczer.
\newblock Detecting and tracking political abuse in social media.
\newblock In {\em Proc. 5th International AAAI Conference on Weblogs and Social
  Media (ICWSM)}, 2011.

\bibitem{Ratkiewicz2011}
J.~Ratkiewicz, M.~Conover, M.~Meiss, B.~Gon\c{c}alves, S.~Patil, A.~Flammini,
  and F.~Menczer.
\newblock Truthy: Mapping the spread of astroturf in microblog streams.
\newblock In {\em Proceedings of the 20th International Conference Companion on
  World Wide Web}, WWW '11, pages 249--252, 2011.

\bibitem{Resnick2014}
P.~Resnick, S.~Carton, S.~Park, Y.~Shen, and N.~Zeffer.
\newblock Rumorlens: A system for analyzing the impact of rumors and
  corrections in social media.
\newblock In {\em Proc. Computational Journalism Conference}, 2014.

\bibitem{Salganik}
M.~J. Salganik, P.~S. Dodds, and D.~J. Watts.
\newblock Experimental study of inequality and unpredictability in an
  artificial cultural market.
\newblock {\em Science}, 311(5762):854--856, 2006.

\bibitem{shao2016hoaxy}
C.~Shao, G.~L. Ciampaglia, A.~Flammini, and F.~Menczer.
\newblock Hoaxy: A platform for tracking online misinformation.
\newblock In {\em Proceedings of the 25th International Conference Companion on
  World Wide Web}, pages 745--750, 2016.

\bibitem{Shao2018anatomy}
C.~Shao, P.-M. Hui, L.~Wang, X.~Jiang, A.~Flammini, F.~Menczer, and G.~L.
  Ciampaglia.
\newblock Anatomy of an online misinformation network.
\newblock {\em PLoS ONE}, 13(4):e0196087, 2018.

\bibitem{shu2017fake}
K.~Shu, A.~Sliva, S.~Wang, J.~Tang, and H.~Liu.
\newblock Fake news detection on social media: A data mining perspective.
\newblock {\em ACM SIGKDD Explorations Newsletter}, 19(1):22--36, 2017.

\bibitem{starbird2017examining}
K.~Starbird.
\newblock Examining the alternative media ecosystem through the production of
  alternative narratives of mass shooting events on twitter.
\newblock In {\em Proceedings of the Eleventh International AAAI Conference on
  Web and Social Media (ICWSM)}, pages 230--239, 2017.

\bibitem{Stroud2011}
N.~Stroud.
\newblock {\em Niche News: The Politics of News Choice}.
\newblock Oxford University Press, 2011.

\bibitem{socialbots-IEEE-DARPA}
V.~Subrahmanian, A.~Azaria, S.~Durst, V.~Kagan, A.~Galstyan, K.~Lerman, L.~Zhu,
  E.~Ferrara, A.~Flammini, F.~Menczer, A.~Stevens, A.~Dekhtyar, S.~Gao,
  T.~Hogg, F.~Kooti, Y.~Liu, O.~Varol, P.~Shiralkar, V.~Vydiswaran, Q.~Mei, and
  T.~Hwang.
\newblock The darpa twitter bot challenge.
\newblock {\em IEEE Computer}, 49(6):38--46, 2016.

\bibitem{Sunstein_extremes}
C.~R. Sunstein.
\newblock {\em Going to Extremes: How Like Minds Unite and Divide}.
\newblock Oxford University Press, 2009.

\bibitem{doi:10.1177/1461444817712086}
C.~J. Vargo, L.~Guo, and M.~A. Amazeen.
\newblock The agenda-setting power of fake news: A big data analysis of the
  online media landscape from 2014 to 2016.
\newblock {\em New Media \& Society}, 20(5):2028--2049, 2018.

\bibitem{botornot_icwsm17}
O.~Varol, E.~Ferrara, C.~A. Davis, F.~Menczer, and A.~Flammini.
\newblock Online human-bot interactions: Detection, estimation, and
  characterization.
\newblock In {\em Proc. Intl. AAAI Conf. on Web and Social Media (ICWSM)},
  2017.

\bibitem{Varol2017epjds}
O.~Varol, E.~Ferrara, F.~Menczer, and A.~Flammini.
\newblock Early detection of promoted campaigns on social media.
\newblock {\em EPJ Data Science}, 6(1):13, 2017.

\bibitem{vonAhn2003}
L.~von Ahn, M.~Blum, N.~J. Hopper, and J.~Langford.
\newblock Captcha: Using hard ai problems for security.
\newblock In E.~Biham, editor, {\em Advances in Cryptology --- Proceedings of
  EUROCRYPT 2003: International Conference on the Theory and Applications of
  Cryptographic Techniques}, pages 294--311. Springer, 2003.

\bibitem{Vosoughi1146}
S.~Vosoughi, D.~Roy, and S.~Aral.
\newblock The spread of true and false news online.
\newblock {\em Science}, 359(6380):1146--1151, 2018.

\bibitem{Wardle_FirstDraft_2017}
C.~Wardle.
\newblock {Fake news. It's complicated.}
\newblock White paper, First Draft News, February 2017.

\bibitem{webb2008social}
S.~Webb, J.~Caverlee, and C.~Pu.
\newblock Social honeypots: Making friends with a spammer near you.
\newblock In {\em Proc. CEAS}, 2008.

\bibitem{FB-INFO-OPS}
J.~Weedon, W.~Nuland, and A.~Stamos.
\newblock Information operations and facebook.
\newblock white paper, Facebook, April 2017.

\bibitem{Pew2018}
S.~Wojcik, S.~Messing, A.~Smith, L.~Rainie, and P.~Hitlin.
\newblock Bots in the twittersphere.
\newblock White paper, Pew Research Center, April 2018.

\bibitem{Howard2017}
S.~C. Woolley and P.~N. Howard.
\newblock Computational propaganda worldwide: Executive summary.
\newblock Working Paper 2017.11, Oxford Internet Institute, 2017.

\bibitem{Wu2018TraceMiner}
L.~Wu and H.~Liu.
\newblock Tracing fake-news footprints: Characterizing social media messages by
  how they propagate.
\newblock In {\em Proc. 11th ACM International Conference on Web Search and
  Data Mining (WSDM)}, 2018.

\bibitem{wu2016mining}
L.~Wu, F.~Morstatter, X.~Hu, and H.~Liu.
\newblock Mining misinformation in social media.
\newblock In {\em Big Data in Complex and Social Networks}, pages 123--152. CRC
  Press, 2016.

\bibitem{Zannettou:2017:WCU:3131365.3131390}
S.~Zannettou, T.~Caulfield, E.~De~Cristofaro, N.~Kourtelris, I.~Leontiadis,
  M.~Sirivianos, G.~Stringhini, and J.~Blackburn.
\newblock The web centipede: Understanding how web communities influence each
  other through the lens of mainstream and alternative news sources.
\newblock In {\em Proceedings of the 2017 Internet Measurement Conference}, IMC
  '17, pages 405--417, 2017.

\bibitem{zubiaga2018detection}
A.~Zubiaga, A.~Aker, K.~Bontcheva, M.~Liakata, and R.~Procter.
\newblock Detection and resolution of rumours in social media: A survey.
\newblock {\em ACM Computing Surveys}, 50, 2018.
\newblock Forthcoming.

\end{thebibliography}

\end{document}